\documentclass[a4paper,10pt]{article}
\usepackage[utf8]{inputenc}
\pdfoutput=1
\usepackage{natbib}
\usepackage{amsfonts}
\usepackage{amsmath}
\usepackage{amsthm}
\usepackage{graphicx}
\usepackage[left=2.2cm, right=2.2cm, top=2.3cm, bottom=2.3cm]{geometry}
\usepackage{caption}
\usepackage{mathrsfs}
\usepackage{setspace}
\usepackage{algorithm}
\usepackage{url}
\usepackage{subfig}
\usepackage{float}
\usepackage{xcolor}

\newtheorem{mylemma}{Lemma}

\title{Bayes Linear Methods for Large-Scale Network Search}
\author{Lisa Turner, Nedialko B. Dimitrov and Paul Fearnhead}

\date{\today}

\definecolor{light-gray}{gray}{0.5}

\newcommand{\ned}[1]{\textcolor{blue}{\textbf{$\mathbf{\langle}$ Ned:} #1 \textbf{$\mathbf{\rangle}$}}}

\newcommand{\lisa}[1]{\textcolor{purple}{\textbf{$\mathbf{\langle}$ Lisa:} #1 \textbf{$\mathbf{\rangle}$}}}
\newcommand{\paul}[1]{\textcolor{purple}{\textbf{$\mathbf{\langle}$ Paul:} #1 \textbf{$\mathbf{\rangle}$}}}

\newcommand{\nedOld}[1]{\textcolor{light-gray}{\textbf{$\mathbf{\langle}$ Ned:} #1 \textbf{$\mathbf{\rangle}$}}}

\begin{document}
 \maketitle

\begin{abstract}
 Consider the problem of searching a large set of items, such as emails, for a small set which are relevant to a given query.
 This can be implemented in a sequential manner -- whereby we use knowledge from earlier items that we have screened to help us
 choose future items in an informed way. Often the items we are searching have an underlying network structure: for example emails
 can be related to a network of participants, where an edge in the network relates to the presence of a communication between 
 those two participants. Recent work by Dimitrov, Kress and Nevo has shown that using the information about the network structure together
 with a modelling assumption that relevant items and participants are likely to cluster together, can greatly increase the rate of screening
 relevant items. However their approach is computationally expensive and thus limited in applicability to small networks. Here we show how Bayes Linear methods
 provide a natural approach to modelling such data; that they output posterior summaries that are most relevant to heuristic policies 
 for choosing future items; and that they can easily be applied to large-scale networks. Both on simulated data, and data from the Enron Corpus, Bayes Linear
 approaches are shown to be applicable to situations where the method of Dimitrov et al. is infeasible; and give substantially better performance
 than methods that ignore the network structure.
\end{abstract}

\section{Introduction}
There are many applications where you wish to search through a large set of items, such as emails or documents
to find a small set of them which are relevant to a query. Often these items are distributed on edges of a network. 
We call the problem of finding relevant items distributed across the edges of a network as \emph{network-based search}.  
Network-based search is increasingly common in many applications and contexts. We give two specific examples, which we return to later in the paper.\\

The first example comes from corporate law suits. Companies are often required to hand over large databases of emails to one another and the information within these emails may make up part of their legal battle \citep{cavaliere2005can}.  A key part Oracle's case in \textit{Oracle America, Inc. v. Google Inc.} included emails suggesting Google were more aware than previously known about how their use of Java APIs could infringe upon licensing agreements \citep{OracleClosing}.  To find and judge which of the emails are relevant to a case requires searching through a 
huge network of emails, the majority of which will be irrelevant, in possibly time pressured situations, at the cost of millions of dollars and thousands of hours \citep{flynn2003mail}.    \\

The second example comes from intelligence processing. Efficient military operations and law enforcement rely on the timely processing of intelligence \citep{hughbank2010intelligence}. 
Overwhelming amounts of intelligence is collected daily, particularly communications intelligence where the use of social media, text messaging and emails have drastically increased \citep{duyvesteyn2014future}. Many terrorist attacks could have been stopped or at least mitigated if the available intelligence was better processed and analysed \citep{gorman2002probe}. One example is the Christmas day bombing of Northwest Airlines flight $253$, for which $14$ intelligence failures were reported \citep{senate2010christmas} including failing to uncover key intelligence reports on the bomber.  The National Intelligence Strategy (2009) states a key area of improvement is to narrow the gap between collecting intelligence and being able to make sense of the intelligence which is collected.  A result of the overwhelming amount of intelligence collected is that often the processor is faced with a bottle neck of intelligence items; far more are collected than can be processed.  
In a time critical situation, the main challenge for the processor is to provide the analysts with the largest network of relevant intelligence items in the given time window.\\

In both these examples, we have a set of emails or other communications, henceforth called \textit{items} for simplicity, and we wish to search the set items to find those which are relevant to the query of interest. Each of the items involve two or more participants.  The participants and the items between them induce a network where the participants form the nodes.  An edge exists between two nodes 
if there is at least one item between the two associated participants. Participants can also be relevant or irrelevant to the query.  Relevant items are more likely to occur between relevant participants.  
In addition, it is likely that relevant participants will cluster in groups.  Thus, the network contains information that can be exploited to help decide which items to observe.  \\

We call the person who searches through the items the \textit{user} and assume correct classification by the user. The user’s task is to identify as many relevant items as possible, where there are far more items than can be observed during a limited time period available.  
Items are observed one at a time, with the user gaining information that can be exploited to focus the future search more effectively.   \\

The problem has three related aspects:
\begin{enumerate}
\item Constructing an appropriate joint prior distribution for the relevance of participants and items. 
\item Updating this joint distribution as items are observed.
\item Deciding which item to observe next given the current joint distribution on the relevance of items and participants. 
\end{enumerate}

\citet{dimitrov2015finding} show that using a prior distribution for the relevance of participants that models the fact that relevant participants cluster together can substantially improve the success of the
user at finding relevant items. Their prior distribution is based upon classifying each participant as either relevant to the query or irrelevant, modelled through a set of unobserved binary random variables, 
one for each participant. The joint distribution of these binary random variables is a specified as a Markov random field (MRF). This Markov random field introduces local dependencies that encourage relevant participants
to cluster together. In addition, associated with each edge is a random variable describing the probability of observing a relevant item between the two associated participants. 
These random probabilities are dependent on the involved participants' relevance values.   An item involving two relevant participants is more likely to 
be relevant to the query then when at least one participant is irrelevant. \\   
 
The limitation of the model of \cite{dimitrov2015finding} is that updating the joint distribution in step (2) can be computationally prohibitive for large networks.   Using an exact inference method such as variable elimination \citep{zhang1994simple}, conditioning \citep{shachter1994global} or 
junction trees \citep{lauritzen1988local}, the computational cost grows exponentially in the tree width of the graph.  
Hence they are intractable for many large networks \citep{koller2009probabilistic}. \citet{ellis2013algorithms} demonstrates 
that the use of exact inference can limit the size of network for which the method is computationally tractable to between $100$ and $700$ nodes,
depending on the connectivity of the network. For larger networks, approximate inference is required to overcome the computational intractability of exact inference. 
\\

There are a number of standard heuristic policies for performing the third step. Importantly for our work, most of these policies depend on the posterior distribution
for the probabilities that items are relevant just through the posterior mean and covariance. This motivates using the Bayes Linear (BL) methodology \citep{goldstein2007bayes}, which summarises the prior and posterior distribution just through their means and covariances, and gives a simple procedure to update these given new data. Thus, the use of BL has many advantages for this application. Firstly, it
simplifies specifying the prior, as we need only specify the prior mean and covariance. Secondly, the BL updates are computationally practical for large networks; with their computational cost scaling,
at worst, as the cube of the number of nodes in the network, compared to exponentially, in the worst case, for the exact Bayesian updates. Finally, whilst the BL updates are approximate, they are approximations that focus on the aspects of the posterior, namely the mean and covariance, that are needed for the decision problem in step 3.\\

Despite these advantages, the standard BL updates are inappropriate for our application, as they ignore the fact that we have a posterior distribution on parameters that represent binary random variables. As such, the expectation of these parameters are constrained to lie in the interval $[0,1]$. The standard BL updates can force the posterior expectation to lie outside this range. We introduce an extension of the BL update that respects this constraint.\\ 

The outline of the chapter is as follows. In Section \ref{secModel}, we introduce the model of \cite{dimitrov2015finding}.  
Section \ref{secBL} gives an overview of BL methodology. In Section \ref{secConstrainedBL} the constrained BL optimisation problem is introduced, where additional constraints are added 
to the BL optimisation problem to ensure the BL posterior means for any probabilities remain in the range $[0,1]$.  We show the constrained BL updates have an analytical solution, allowing for savings in the 
computational cost of inference compared with using the constrained optimisation problem directly.  Section \ref{secBLintelligence} describes how the BL updates are used for approximate 
inference within the network based search method. We evaluate the accuracy of the BL procedure both for approximating the posterior distribution of the Markov random field model, and 
for the network based search problem in Section \ref{secResults}. These results include analysis of data based on a terrorist network and data taken from the Enron Corpus. In particular we show how BL can perform
network based search for large-scale networks, and that it leads to substantially improved performance over methods that ignore the dependence structure implied by the network.

\section{The Model}
\label{secModel}
The model we use for intelligence collection is originally described by \cite{nevo2011information}.  Let $G=(V,E)$ denote the graph associated with the network.  
The nodes in the graph, $V$, represent the participants.  The edge $(u,v) \in E$ exists if and only if there is at least one item between participants $u$ and $v$.  \\

Let there be $m$ nodes in the graph and $n$ edges. Associated with each node is a random variable describing how relevant the participant is to the query.  
Define this random variable as $Z_u$ for $u \in 1,\ldots,m$ and the set of participants' relevance values by $\mathbf{Z}=\left\{ Z_1,\ldots,Z_{m} \right\}$.  
We model the relevance of participant $u$ as a binary random variable where:
\begin{align*}
 Z_u = \left\{
  \begin{array}{ll}
    1 & \text{if participant $u$ relevant to query}\\
    0 & \text{otherwise}
  \end{array}
\right. .
\end{align*}
A random variable $P_{uv} \mid Z_{u},Z_{v}$ for edge $(u,v)\in E$, describes the probability of observing a relevant item on that edge given $Z_u$ and $Z_v$.  
Let the set of probabilities be $\mathbf{P} = \left\{ P_{uv} , (u,v) \in E \right\}$.  Given the involved participants' relevance values, the probabilities are conditionally independent. 
For conjugacy, we model these prior conditional probabilities using a beta distribution,
\begin{align}
 P_{uv}\mid Z_u,Z_v \sim \text{Beta} (a(Z_u,Z_v),b(Z_u,Z_v)).
\end{align}

When an item is observed, it is classed as relevant or irrelevant to the query. Suppose the $k$th item is observed on edge $(u,v)$, and $Y_{uv}^k$ denotes the observed relevance value of the item.  
That relevance value is modelled by the Bernoulli distribution with success probability $P_{uv}$.
After $n_{uv}$ observations on edge $(u,v)$, the outcome of the observed items is stored through the sufficient statistics $n_{uv}$ and $Y_{uv}$ where, \begin{align}
 Y_{uv} = \sum_{k = 1}^{n_{uv}} Y_{uv}^k, \label{suffstat1}
\end{align}
is the number of relevant items observed.  The posterior joint distribution over the participants' relevance values and the probabilities that edges produce relevant items, given the sufficient statistic
$Y_{uv}$ on each edge, $\mathbf{Y} = \left\{ Y_{uv},(u,v)\in E \right\}$ is:

 \begin{align}
P(\mathbf{Z},\mathbf{P}\mid \mathbf{Y}) & = P(\mathbf{Z}\mid\mathbf{Y}) P( \mathbf{P}  \mid \mathbf{Z}, \mathbf{Y}), \nonumber \\
&=  P(\mathbf{Z}\mid \mathbf{Y}) \left[\prod_{(u,v) \in E} P(P_{uv}\mid Z_{u},Z_{v}, Y_{uv}) \right]. \label{joint_discrete}
 \end{align}
 The equality above comes from the fact that given values of $Z_{u},Z_{v}$, the variable $P_{uv}$ is independent of other $P_{u'v'}$, $Y_{u'v'}$ variables.  The conditional posterior for each $P_{uv}$ will be
 \[
  P_{uv}\mid Z_{u},Z_{v},Y_{uv} \sim \mbox{Beta}(a(Z_u,Z_v)+Y_{uv},b(Z_u,Z_v)+n_{uv}-Y_{uv}).
 \]
 
 \cite{dimitrov2015finding} complete this model by specifying a Markov Random Field model for $\mathbf{Z}$. Such a model introduces local dependence structure with, for example, participants that share an edge in the 
 network being more likely to be of the same type: either both relevant or both irrelevant. Whilst a natural model, this leads to difficulties with evaluating this joint posterior distribution as 
 calculating $P(\mathbf{Z}\mid \mathbf{Y})$ can be computationally prohibitive.  Using exact inference algorithms the computational cost of evaluating this grows exponentially, 
 in the worst case, with the tree width of the graphical model \citep{koller2009probabilistic}.  See \citet{nevo2011information} and \citet{ellis2013algorithms} for more information on the exact updating process.

\subsection{Sequential Decision Making}
The reason we need to calculate the posterior distribution \eqref{joint_discrete}, is that given the current set of observations, $\mathbf{Y}$, we want to choose which item to observe next.  We can define this as a Bayesian sequential decision problem, where we wish to maximise the number of relevant items observed over a fixed time interval, with the policy of which item to observe next depending on which of the items to date have been relevant.  Solving this decision problem optimally is intractable, but there are many heuristic policies that have been shown to perform well for such decision problems \citep{auer2002finite}.  \\

A simple policy would be to observe the item which you currently think is most likely to be relevant.  That is, we would choose an item from the edge $(u,v)$ for which the posterior expectation of $P_{uv}$ is highest.  This is called the greedy policy.  In practice, the greedy policy can often perform poorly, particularly for decision problems over a long time interval.  It just exploits the current information as opposed to also trying to learn more about which edges have the highest $P_{uv}$ values.  As a result, there are more refined heuristic methods, that take account of not just the posterior means of the $\mathbf{P}$ but also the posterior variances \citep{lai1987adaptive, kaelbling1993learning, may2012optimistic}.  Informally, these choose edges which do not just have higher means but also higher variances. \\

Thus to implement one of these policies, we do not need to calculate the full posterior distribution for the $P_{uv}$'s, but just the posterior mean and variance. This motivates the use 
of Bayes linear methods, which are based on Bayesian modelling and updating that solely use the mean and variance. 

\section{Bayes Linear}
\label{secBL}
Bayes linear \citep{goldstein2007bayes} replaces the exact Bayesian update in \eqref{joint_discrete} with an approximation.  The idea of BL is to consider only the mean and covariance of the parameters. This simplifies the prior specification as only a mean and variance, rather than the full distribution, is needed.  On observing data, these are updated to produce approximations to the
posterior mean and variance. For our application, BL requires specifying the prior expectation of the latent variables, $\mathbf{Z} = \left\{ Z_1,\ldots,Z_m \right\}$, and the observable quantities $\mathbf{Y} = \left\{ Y_1,\ldots,Y_n \right\}$.  The uncertainty in the expectations of the random variables, and the extent that one random variable will influence another, is specified through
\begin{align*}
 \text{Cov}(\mathbf{Z},\mathbf{Y}) = \begin{pmatrix} \text{Var}(\mathbf{Z}) & \text{Cov}(\mathbf{Z},\mathbf{Y}) \\ \text{Cov}(\mathbf{Y},\mathbf{Z}) & \text{Var}(\mathbf{Y}) \end{pmatrix},
\end{align*}
where: $\text{Var}(\mathbf{Z})$ is the covariance matrix for the latent variables, describing how they linearly interact; $\text{Var}(\mathbf{Y})$ is the covariance matrix describing how the observations linearly interact; and $\text{Cov}(\mathbf{Y},\mathbf{X})$ is the covariance matrix describing how an observation in $\mathbf{Y}$ will linearly influence beliefs in the latent variables. \\

The BL updates can be defined in terms of finding the best estimate of each $Z_k$ by a linear combination of the data.  This best estimate is defined in terms of minimising mean squared error.  The BL posterior mean for $Z_k$ is just the resulting estimate and the BL posterior variance is defined as the variance of these estimators.  Formally, for each $k \in \left\{1,\ldots,m\right\}$ we wish to find the coefficients $\mathbf{h}^k = \left( h_0^k,h_1^k,\ldots,h_n^k \right)$ that solve the following optimisation problem:
\begin{equation}
\begin{aligned}
& \underset{\mathbf{h}^k}{\text{minimise}} \ E \left[ \left( Z_k - h_0^k -\sum_{i=1}^n h_i^k Y_i  \right)^2 \right] , \label{standardBLopt} & k \in \{1,\ldots,m\}.
\end{aligned}
\end{equation}
Then we define the estimated posterior expectation as:
\begin{align}
    \widehat{E} \left[ Z_k \mid \mathbf{Y} \right]  =h_0^k + \sum_{i=1}^n h_i^k Y_i, && k \in \{1,\ldots,m\}, \label{unconst_BLE_opt}
\end{align}
and the updated estimated posterior covariance between $Z_k$ and $Z_l$ as:
\begin{align}
 \widehat{\text{Cov}} \left( Z_k , Z_l \mid \mathbf{Y} \right) =  E \left[ \left( Z_k - h_0^k - \sum_{i=1}^n h_i^k Y_i  \right) \left( Z_l - h_0^j - \sum_{j=1}^n h_j^l Y_j \right) \right] , && k,l \in \{1,\ldots,m\}. \label{unconst_BLCov_opt}
\end{align}
The expectation in equation \eqref{standardBLopt} is over both the latent variables and the observable quantities. By multiplying out the expectation, we can see that $\mathbf{h}^k$ depends on the prior specification of the expectation and covariance of $\mathbf{Z}$ and $\mathbf{Y}$.   The optimisation problems, in \eqref{standardBLopt}, are standard convex quadratic optimisation problems \citep{boyd2004convex} and can be solved analytically to give

\begin{align}
 \mathbf{h}_{1:n}^k = \text{Cov}(Z_k,\mathbf{Y})\text{Var}(\mathbf{Y})^{-1},
\end{align}
and $h_{0}^k = E[Z_k] - (\mathbf{h}_{1:n}^k)^T \mathbf{y}$.  Thus, the BL updated expectation is
\begin{align}
 \widehat{E} \left[ Z_k\mid\mathbf{Y} \right] &= E\left[Z_k \right] + \text{Cov}(Z_k,\mathbf{Y}) \text{Var}(\mathbf{Y})^{-1}(\mathbf{y} - E\left[ \mathbf{Y} \right]), \label{EZ_BL}
\end{align}
and BL updated covariance is
\begin{align}
  \widehat{\text{Cov}} \left( Z_k,Z_l \mid \mathbf{Y} \right) &= \text{Cov}(Z_k,Z_l) - \text{Cov}(Z_k,\mathbf{Y})  \text{Var}(\mathbf{Y})^{-1} \text{Cov}(\mathbf{Y},Z_l). \label{CovZ_BL}
\end{align}
For a full derivation of the update equations see \cite{goldstein2007bayes}.

\subsection{Constrained Bayes Linear}
\label{secConstrainedBL}
For a set of binary latent variables, we have the property that the true posterior expectation of the latent variable $Z_u$, $u \in \left\{ 1,\ldots,m\right\}$ will be in the range $[0,1]$.  
Therefore, a desirable property of any approximation to the posterior expectation is that this still holds.  The BL updated expectation, \eqref{EZ_BL}, does not necessarily 
have this property. \\

To overcome this, we can recast the BL updates in terms of their original optimisation problem and modify \eqref{standardBLopt} to include appropriate constraints on the posterior mean.  
In our case, the desirable property is that the updated expectation of the random variables remains in the range $[0,1]$. This can be achieved by adding linear inequality constraints to the BL optimisation problem. The constrained form of BL updates, for binary random variables, is given by:
\begin{equation}
\begin{aligned}
& \underset{\mathbf{h}^k}{\text{minimise}} \label{constrainedBLopt}
& & E \left[ \left( Z_k - h_0^k - \sum_{i=1}^n h_i^k Y_i  \right)^2 \right] ,\\
& \text{subject to}
& & h_0^k + \sum_{i=1}^n h_i^k y_i \leq 1, \\
& & & h_0^k + \sum_{i=1}^n h_i^k y_i \geq 0,
\end{aligned}
\end{equation}
where $\mathbf{y} = (y_1,\ldots, y_n)$ are the observed quantities. The values needed for the constrained optimisation problem \eqref{constrainedBLopt} are found by expanding the objective function
\begin{align}
 E \left[ \left( Z_k -h_0^k - \sum_{i=1}^n h_i^k Y_i  \right)^2 \right] = E[Z_k^2] &-2h_0^k E[Z_k] -  2 \sum_{i=1}^n h_i^k E[Z_kY_i] + 2h_0^k \sum_{i=1}^n h_i^k E[Y_i] \nonumber \\
 &+ (h_0^k)^2 + \sum_{i=1}^n\sum_{j=1}^n h_i^k  h_j^k E[Y_iY_j].
\end{align}
As in the case of unconstrained BL updates, the resulting optimisation problem is convex because the coefficients of the square terms, $h_i^k h_j^k$, make a positive semi-definite matrix.  Because of this, the constrained optimisation problem can be solved using available convex optimisation software \citep{andersen2013cvxopt}.  Constrained BL updates produce fundamentally different solutions than unconstrained BL updates.  For example, for unconstrained BL updates, the $\mathbf{h}^k$ do not depend on the actual value of the observations, just their expectation and covariances.  On the other hand, when one of the constraints in \eqref{constrainedBLopt} is binding, $\mathbf{h}^k$ will depend on the observations.  Once $\mathbf{h}^k$ values are computed, BL updated expectations and covariances are found using equations \eqref{unconst_BLE_opt} and 
\eqref{unconst_BLCov_opt}.  \\

Solving problem \eqref{constrainedBLopt} using a convex optimisation solver can be slow in practice.  However, it is possible to derive a fast algorithm for problem \eqref{constrainedBLopt} through analytical solutions to related equality constrained quadratic programs using Lemma \ref{tightconst}.

\begin{mylemma} \label{tightconst} 
If the BL updated expectation $\hat{E}[Z_k|\mathbf{Y}]$ for the unconstrained problem, \eqref{standardBLopt}, is between $[0,1]$, the solution to problem \eqref{constrainedBLopt} is the same as that of the unconstrained problem, \eqref{standardBLopt}.  Otherwise, one of the constraints to problem \eqref{constrainedBLopt} is tight.  If $\hat{E}[Z_k|\mathbf{Y}] > 1$, the constraint $h_0^k + \sum_{i=1}^n h_i^k y_i \leq 1$ is tight and if $\hat{E}[Z_k|\mathbf{Y}]  < 0$, the constraint $h_0^k + \sum_{i=1}^n h_i^k y_i \geq 0$ is tight.
\end{mylemma}
The proof for Lemma \ref{tightconst} is given in Appendix \ref{proof_lemma_tightconst}. This allows us to motivate the following method for solving the constrained BL optimisation.  Solving the unconstrained BL update, through \eqref{EZ_BL} and \eqref{CovZ_BL}, allows us to identify which constraint, if any, in \eqref{constrainedBLopt} is tight.  Once the tight constraint is identified, we can derive an analytical solution to the corresponding equality constrained problem.  The benefit in computational time comes from the fact that solving \eqref{constrainedBLopt} is reduced to several matrix multiplications, as opposed to using repeated gradient descent type methods required for general convex optimisation.  More specifically, the algorithm to solve \eqref{constrainedBLopt} for each $k=1,\ldots,m$, is as follows.
\begin{enumerate}
    \item Solve for the unconstrained BL update for the expectation through \eqref{EZ_BL}.
    \item If $\widehat{E} \left[ Z_k\mid\mathbf{Y} \right] $ is between $[0,1]$, the solution to problem \eqref{constrainedBLopt} is the same as that of the unconstrained BL update.
    \item Otherwise, one of the constraints of problem \eqref{constrainedBLopt} is tight in an optimal solution. To compute the optimal solution, 
solve:
\begin{equation}
\begin{aligned}
& \underset{\mathbf{h}^k}{\text{minimise}} \label{constrainedBLopt_equalc}
& & E \left[ \left( Z_k -h_0^k - \sum_{i=1}^n h_i^k Y_i  \right)^2 \right] ,\\
& \text{subject to}
& &h_0^k + \sum_{i=1}^n h_i^k y_i = c,
\end{aligned}
\end{equation}
with $c = 0$ if $\widehat{E} \left[ Z_k\mid\mathbf{Y} \right] <0$ and $c=1$ if $\widehat{E} \left[ Z_k\mid\mathbf{Y} \right] >1$. 
\end{enumerate}
To solve problem \eqref{constrainedBLopt_equalc} quickly, we can make use of the following lemma.  
\begin{mylemma} \label{lemma_equalc_const}
 The analytical solution to the optimisation problem \eqref{constrainedBLopt_equalc} is:
 \begin{align}
 (\mathbf{h}_{1:n}^k)^T &= \left(\text{Cov}(Z_k,\mathbf{Y}) + (c-E[Z_k])(\mathbf{y}- E[\mathbf{Y}])^T\right)\left( \text{Var}(\mathbf{Y}) + (\mathbf{y} - E[\mathbf{Y}])(\mathbf{y}-E[\mathbf{Y}])^T\right)^{-1},  \label{eq:l1}
  \end{align}
  where $\mathbf{h}_{1:n}^k = (h_1^k,\ldots,h_n^k)^T$, and:
  \begin{align}
h_0^k &= c - (\mathbf{h}_{1:n}^k)^T \mathbf{y}. \nonumber
 \end{align}
\end{mylemma}
The proof of Lemma \ref{lemma_equalc_const} can be found in the Appendix \ref{proof_lemma_equalc_const}.  The updated expectation and covariance can then by calculated from equations \eqref{unconst_BLE_opt} and \eqref{unconst_BLCov_opt}.  These analytical solutions provide a fast method for computing constrained BL updates.\\

The computational cost of both constrained BL and unconstrained BL, for a set of observations $\mathbf{Y} = \left\{ Y_1,\ldots,Y_n \right\}$ and set of latent variables $\mathbf{Z} = \left\{ Z_1,\ldots,Z_m \right\}$ is $O(m^2n+n^2m)$.  This comes from the cost of solving the system of linear equations for $(h_{1:n}^k)^T$ which takes $O(n^2)$ for each of the $k=1,\ldots,m$ latent variables giving a cost of $O(mn^2)$ and calculating the updated covariance at a cost of $O(m^2n)$.

\section{Bayes Linear for the Network Based Searches}
\label{secBLintelligence}
The computational bottleneck of exact inference for the process is updating the beliefs on the participant's relevance values, $\mathbf{Z}$.  
We apply constrained BL updates to approximate the posterior mean and covariance of these latent variables.  The constrained BL method for network based searches is given in Algorithm \ref{Alg_BLsearch}.
\begin{algorithm}
\caption{Bayes Linear Network-Based Search}
\label{Alg_BLsearch}
\begin{enumerate}
 \item Calculate approximations to quantities required for BL updates given the current set of observations. See Section \ref{BLintelligence_PriorValues}. 
\item Find $\widehat{E}[\mathbf{Z}\mid\mathbf{Y}]$ and $\widehat{\text{Var}}(\mathbf{Z}\mid\mathbf{Y})$ using constrained BL updates. 
\item Calculate $\widehat{E}[\mathbf{P}\mid\mathbf{Y}]$ and $\widehat{\text{Var}}(\mathbf{P}\mid\mathbf{Y})$. See Section \ref{sec_approx_joint}.
\item Decide which item to observe next, using $\widehat{E}[\mathbf{P}\mid\mathbf{Y}]$ and $\widehat{\text{Var}}(\mathbf{P}\mid\mathbf{Y})$.  
\end{enumerate}
\end{algorithm}

\subsection{Approximating the Bayes Linear Prior Values from  $E[\mathbf{Z}]$, $\text{Var}(\mathbf{Z})$ and $\mathbf{P}|\mathbf{Z}$} \label{BLintelligence_PriorValues}
We assume the prior expectation and variance of the participants' relevance values are given, along with a prior conditional beta distribution for $\mathbf{P} \mid \mathbf{Z}$.  Based on Section \ref{secBL}, performing BL updates also requires $E[\mathbf{Y}]$, $E[\mathbf{Y}\mathbf{Y}^T]$, $E\left[\mathbf{Z} \mathbf{Y}\right]$ and $E\left[\mathbf{Z}\mathbf{Z}^T\right]$.  The value of $E\left[\mathbf{Z}\mathbf{Z}^T\right]$ can be calculated directly from the priors given to the user; the remaining quantities must be approximated from $E[\mathbf{Z}\mathbf{P}]$, $E[\mathbf{P}]$ and $E[\mathbf{P}\mathbf{P}^T]$, see Section \ref{sec_approx_joint}.  Furthermore, once the updated expectation and covariance of the $\mathbf{Z} \mid \mathbf{Y}$'s are found, this method is used to approximate the updates for $\mathbf{P} \mid \mathbf{Y}$.    \\

The observable quantities, $\mathbf{Y}$, are the number of relevant observations on each edge.  The prior mean and covariance depends on the number of observations on each edge and the prior mean and covariance of the probabilities for each edge. See Lemma \ref{lemmaprior} for the analytical solutions.

\begin{mylemma} \label{lemmaprior}  
The expectations  $E[\mathbf{Y}]$, $E[\mathbf{Y}\mathbf{Y}^T]$ and $E\left[\mathbf{Z} \mathbf{Y}\right]$ can be calculated analytically from $E[\mathbf{Z}]$, $\text{Var}(\mathbf{Z})$ and the prior conditional distribution for $\mathbf{P} \mid \mathbf{Z}$.  The analytical solution for $E[\mathbf{Y}]$ and $E[\mathbf{ZY}]$ can be calculated from
\begin{align}
 E \left[ Y_{uv}  \right] &=n_{uv} E\left[P_{uv} \right],
\end{align}
and:
\begin{align}
 E \left[ Z_k Y_{uv}  \right] = n_{uv} E\left[ Z_kP_{uv} \right]. 
\end{align}
where $n_{uv}$ are the number of observations on edge $(u,v)$ to date.  For $E[\mathbf{Y}\mathbf{Y}]$, the diagonal entries are:
\begin{align}
 E\left[Y_{uv}^2\right] = n_{uv}(n_{uv} - 1) E \left[ \text{Var}(P_{uv}\mid Z_u,Z_v) + E\left[ P_{uv} \mid Z_u,Z_v \right]^2 \right]+ n_{uv}^2 E \left[P_{uv} \right],
\end{align}
whilst the off diagonal entries are given by:
\begin{align}
 E \left[ Y_{uv}Y_{ij} \right] = n_{uv}n_{ij} E[P_{uv}P_{ij}].
\end{align}

\end{mylemma}
The proofs of Lemma \ref{lemmaprior} is in Appendix \ref{prooflemmaprior}.

\subsubsection{Approximating Joint Distributions of $\mathbf{Z}$'s from Expectations and Variances.} \label{sec_approx_joint}

Several of the solutions in Lemma \ref{lemmaprior} require the joint distribution over several participants' relevance values and probabilities.  However, we only have a prior mean and covariance of $\mathbf{Z}$ and the joint distribution for more than two binary $Z$ values is not uniquely defined by their expectation and covariance.  For a set of latent variables, we can approximate a possible joint distribution from the expectation and covariance matrix using BL updates. This joint distribution will have a mean and covariance which match those used to simulate it. For a set of random variables $\mathbf{Z} = (Z_1,Z_2,\ldots,Z_k)$, we assign a realisation of the joint distribution, $\mathbf{z^{(i)}} = (z_1^{(i)},z_2^{(i)},\ldots,z_k^{(i)})$, a probability proportional to
\begin{align}
 \tilde{P}(\mathbf{z}^{(i)}) = \tilde{P}(z_1^{(i)})P(z_2^{(i)}\mid z_1^{(i)}) \ldots \tilde{P}(z_k^{(i)}\mid z_i^{(i)},\ldots,z_{k-1}^{(i)})
\end{align}
where:
\begin{align}
 \tilde{P}(z_j^{(i)}|z_{1}^{(i)},\ldots,z_{j-1}^{(i)}) =  \hat{E}[Z_j| z_{1}^{(i)},\ldots,z_{j-1}^{(i)}]^{z_j^{(i)}} (1 -  \hat{E}[Z_j| z_{1}^{(i)},\ldots,z_{j-1}^{(i)}])^{1-z_j^{(i)}}
\end{align}
and $\hat{E}[Z_j| z_{1}^{(i)},\ldots,z_{j-1}^{(i)}]$ is the updated expectation found using the BL approximation.\\

From this approximated distribution and the conditional beta distributions, we can calculate the required expectations.  We describe here the calculations for $E[P_{uv}P_{ij}]$. 
\begin{align*}
E[P_{uv}P_{ij}] &= E\left[ E\left[P_{uv}P_{ij}|Z_u,Z_v,Z_i,Z_j\right] \right] \\
		&= \sum_{z_u,z_v,z_i,z_j \in \left\{0,1\right\}} \tilde{P}(z_u,z_v,z_i,z_j)E[P_{uv}|z_u,z_v]E[P_{ij}|z_i,z_j],
\end{align*}
as $E\left[P_{uv}P_{ij}|Z_u,Z_v,Z_i,Z_j\right]  = E[P_{uv}|z_u,z_v]E[P_{ij}|z_i,z_j]$ by conditional independence. \\

Similar calculations are use for $E\left[ Z_kP_{uv} \right]$, $E \left[\text{Var}(P_{uv}\mid Z_u,Z_v)\right]$ and $E\left[ E\left[ P_{uv} \mid Z_u,Z_v \right]^2 \right]$, 
and these are given in Appendix \ref{approx_priorvals}.  These quantities need to be calculated for steps 1 and 3 of Algorithm \ref{Alg_BLsearch}. The calculation for step 1 use the prior distributions, and thus can be carried out just once regardless of the number of items processed. For step 3 we need to use the posterior mean and covariance for $\mathbf{Z}$ to obtain the BL estimate of the posterior mean and variance for $\mathbf{P}$. This calculation would need to be repeated prior to choosing each item.

\section{Results}
\label{secResults}
In this section, we analyse the BL model for the network based search process.  Firstly, for simple networks, where observations are simulated from a binary MRF model, 
we consider the errors induced using the BL model compared to the binary MRF model for the updating process, and we consider the performance of the decision problem. 
We empirically show these errors are small in the context of the sequential decision problem, see Section \ref{BLvsExact}.  
In particular, in Section \ref{SecTanzaniaDecision}, we show there is little difference in the performance of the sequential decision problem with the BL model compared with the MRF model, even when though the MRF
is used to simulate the data.\\

Secondly, we compare the BL model to a simple model which assumes independence between each edge, for a set of networks where the underlying model is not simulated from a binary MRF.  We show that even when there is little correlation in the network, the BL model is not detrimental to the performance of the decision algorithm.  When the networks are correlated, the BL model results 
in a higher number of relevant items being observed.  Finally, in Section \ref{SecEnronDecision}, we show the BL model gives good performance for communication networks, which are a subset of the Enron Corpus database.

\subsection{Bayes Linear as an Approximation to the Binary MRF Model} \label{BLvsExact}
To evaluate the accuracy of the BL models as an approximation to the binary MRF model, 
we assume the true underlying model for the networks is a binary MRF model. 
We consider a model where the joint probability of $\mathbf{Z}$ is proportional to the product of a set of factors, with each factor associated with an edge in the network. We use factors of the form
in Table \ref{prior_factor}.  For $\lambda_{i} > 0 $ ($i = 1,2$) 
the network exhibits the property of homophily: the binary random variables on nodes that are connected by an edge are more likely to be of the same value. The large $\lambda_{1}$ and $\lambda_{2}$ are, 
the more likely that such random variables will both be 0 or both be 1 respectively.  We model the conditional probability of observing a relevant item on an edge, given the involved node relevancies as a beta distribution.  Two conditional prior beta distributions are considered to define $\mathbf{P} \mid \mathbf{Z}$; Tables \ref{prior_prob1} and \ref{prior_prob2} for prior conditional 1 and 
prior conditional 2 respectively.  Prior conditional 1 is more skewed to the belief that we are less likely to observe relevant items between participants, even when both participants are considered relevant to the query. \\

\begin{table}[h]
 \centering
  \subfloat[Prior Clique Factor]{
  \begin{tabular}[b]{|cc|c|}\hline
   $Z_i$ & $Z_j$ & $\phi(Z_i,Z_j)$ \\ \hline 
    $0$ & $0$ & $1+\lambda_1$  \\
    $0$ & $1$ & $1$  \\
    $1$ & $0$ & $1$  \\
    $1$ & $1$ & $1+\lambda_2$ \\ \hline
  \end{tabular}
\label{prior_factor}}
\hspace{10pt}
     \subfloat[Prior Conditional 1]{
  \begin{tabular}[b]{|cc|cc|}\hline
   $Z_i$ & $Z_j$ & $a(Z_i,Z_j)$ & $b(Z_i,Z_j)$ \\ \hline
    $0$ & $0$ & $1$ & $9$  \\
    $0$ & $1$ & $1$ & $4$ \\
    $1$ & $0$ & $1$ & $4$  \\
    $1$ & $1$ & $1$ & $1$\\ \hline
  \end{tabular}
\label{prior_prob1}} \vspace{10pt}
     \subfloat[Prior Conditional 2]{
  \begin{tabular}[b]{|cc|cc|}\hline
   $Z_i$ & $Z_j$ & $a(Z_i,Z_j)$ & $b(Z_i,Z_j)$ \\ \hline
    $0$ & $0$ & $1$ & $9$  \\
    $0$ & $1$ & $1$ & $4$ \\
    $1$ & $0$ & $1$ & $4$  \\
    $1$ & $1$ & $9$ & $1$\\ \hline
  \end{tabular}
\label{prior_prob2}}
\caption{\small{ The prior distributions used to define the binary MRF model.  Table \ref{prior_factor} gives form of the prior clique factor used to define the prior MRF model for the participants. Table \ref{prior_prob1} and \ref{prior_prob2} give the two parameters of the conditional beta distributions used to define $\mathbf{P}|\mathbf{Z}$}  } 
\end{table}

The prior mean and covariance required for the BL model are set equal that of the binary MRF model. The BL updated expectations and variances are calculated for a sequence of observations. These updated expectations and variances are compared to the corresponding values in the binary MRF model updated using exact inference methods.  We assume that if the BL updated expectations and covariances are close to these values, the BL model provides a good approximation to the binary MRF model. 

\subsubsection{Simple Line Network} \label{SecSimpleNetwork}
\begin{figure}[h]
 \centering
 \subfloat[Updated values using the BL model and binary MRF model with set of observations $\mathbf{Y} = (Y_{01}^1 = 0, Y_{12}^1 = 0, Y_{12}^2 = 1,Y_{12}^3 = 0, Y_{01}^2 = 0, Y_{12}^4 = 1, Y_{01}^3 = 1)$]{ \centering
    \includegraphics[width = \textwidth]{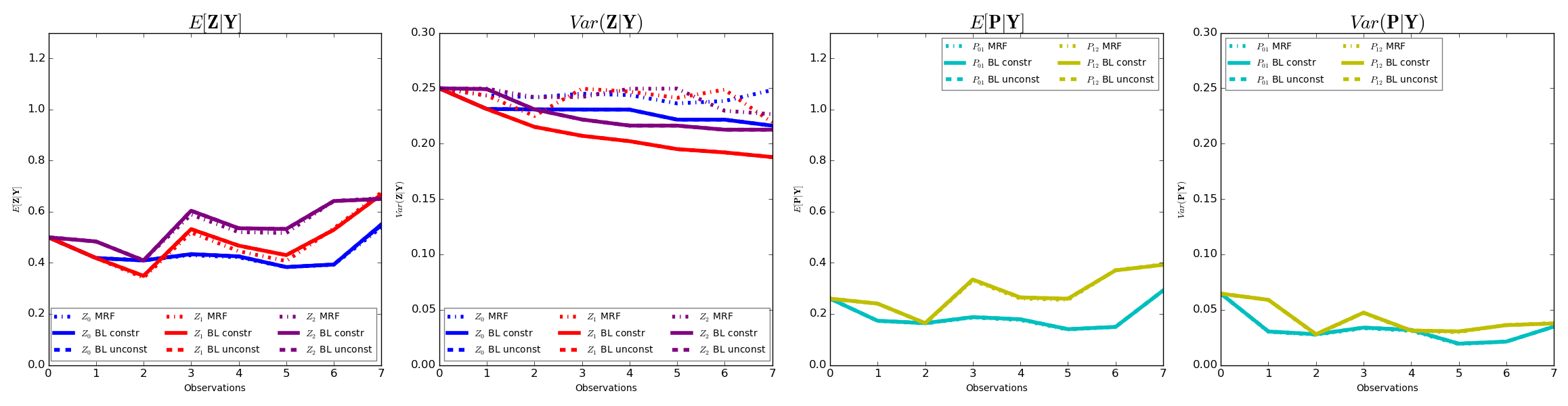}
    \label{noneed_const} }\\
 \subfloat[Updated values using the BL model and binary MRF model  with set of observations $\mathbf{Y} = (Y_{01}^1 = 1, Y_{12}^1 = 1, Y_{12}^2 = 1,Y_{12}^3 = 1, Y_{01}^2 = 1, Y_{12}^4 = 1, Y_{01}^3 = 1)$]{ \centering
    \includegraphics[width = \textwidth]{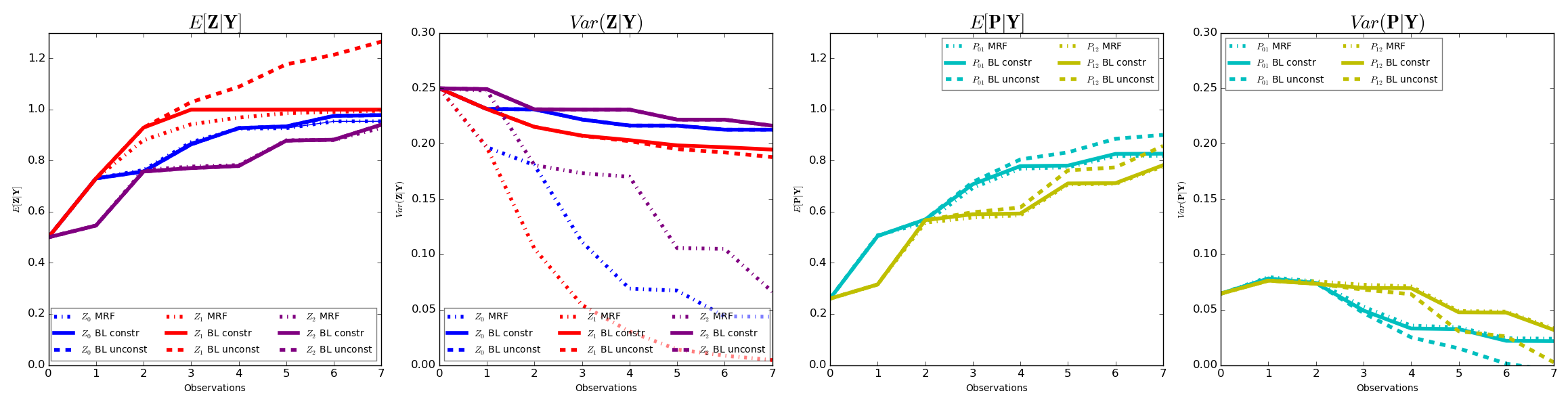}
    \label{need_const} }
  \caption{\small{ The updated expectations and variances of the participants' relevance values, $\mathbf{Z}$, and probabilities, $\mathbf{P}$, using the MRF, unconstrained BL, and constrained BL models given two sets of observations.  The prior clique factor is given in Table \ref{prior_factor} with $[\lambda_1,\lambda_2]  = [0.5,0.5]$, and prior conditional probability distribution in Table \ref{prior_prob1}.  The prior values for BL models are calculated directly from the binary MRF prior model.}  }   
  \label{linenetworkeg}
\end{figure}

We first simulated data for a simple line network with three nodes and two edges. The edges connect $Z_1$ with $Z_2$ and $Z_2$ with $Z_3$. 
The binary MRF model is defined using the prior clique factor in Table \ref{prior_factor} with $[\lambda_1,\lambda_2] = [0.5,0.5]$ and using prior conditional 1 (Table \ref{prior_prob1}).  Figure \ref{linenetworkeg} shows the updated expectations and variances for both $\mathbf{Z} \mid \mathbf{Y}$ and $\mathbf{P}\mid \mathbf{Y}$ using constrained BL model, unconstrained BL model and the binary MRF model for two sets of observations. The unconstrained BL model and the constrained BL model give the same updated values for the set of observations used in Figure \ref{noneed_const}.  These values remain close to the values from the binary MRF model apart from the $\text{Var}(\mathbf{Z}|\mathbf{Y})$.  However, for binary random variables, the mean determines the variance of the random variable, so these values are somewhat redundant. \\

Figure \ref{need_const} shows a set of observations for which the constrained updates ensure the updated expectation of $\mathbf{Z}|\mathbf{Y}$ remains in the correct range. As well as giving a more mathematically elegant solution, the benefit of constrained BL model can be seen in the updated expectation and variance of $\mathbf{P}|\mathbf{Y}$ where the constrained BL updates 
are closer to the updates in the binary MRF then when the unconstrained updates are used.

\subsubsection{Tanzania Network} \label{SecTanzaniaNetwork}
\begin{figure}[h]
 \centering
 \subfloat[Terrorist network]{ \centering
    \includegraphics[width = 0.3\textwidth]{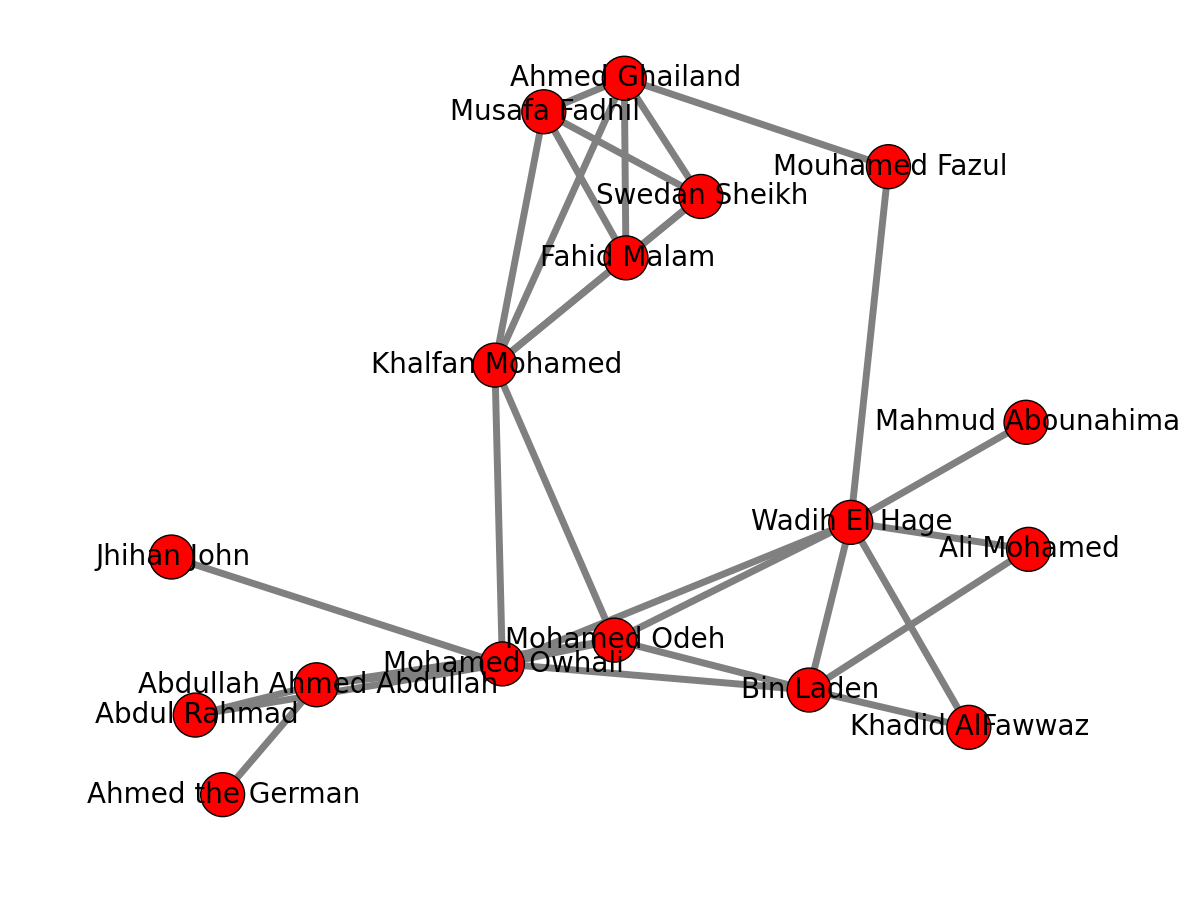}
    \label{Tanzania_terrorists} } \hspace{3pt}
    \subfloat[Network: Prior Conditional 1]{ \centering
    \includegraphics[width = 0.3\textwidth]{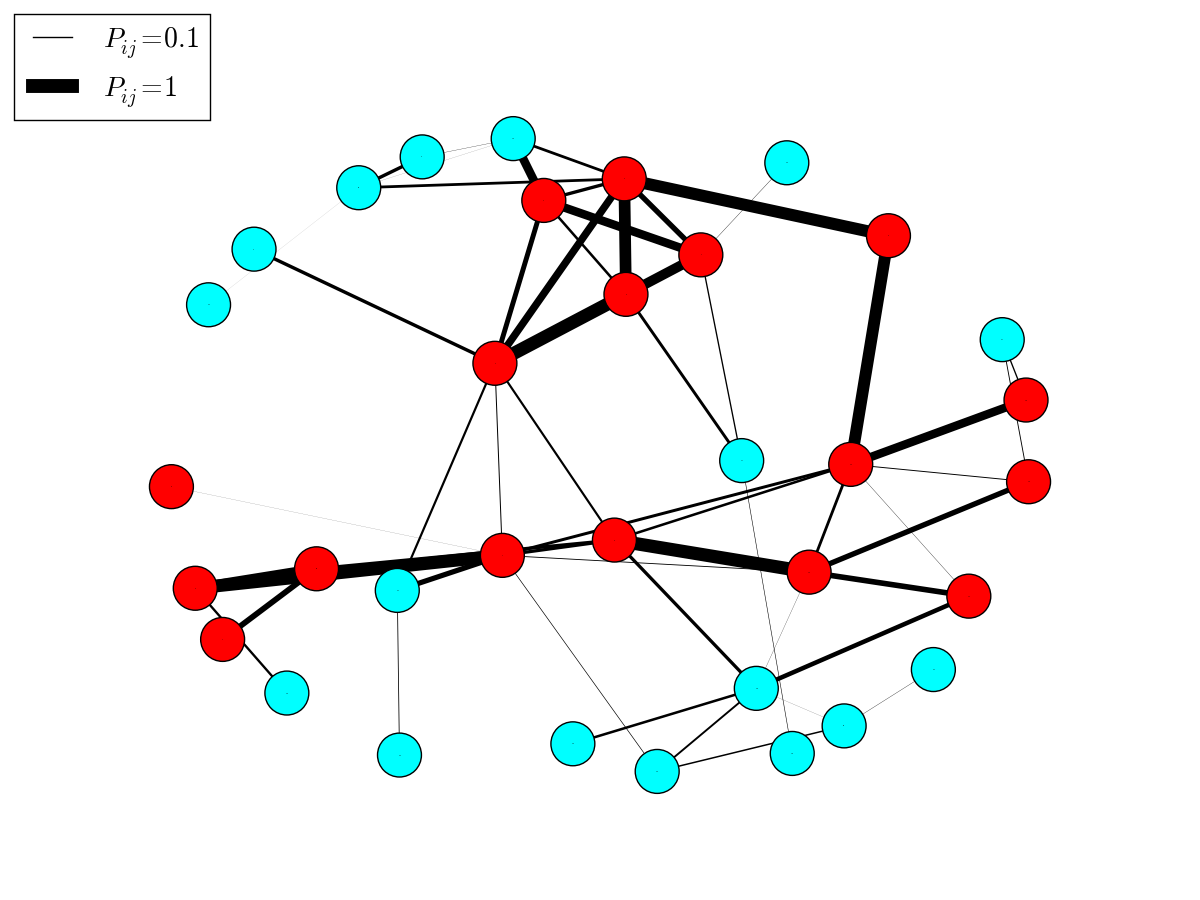}
    \label{TanzaniaNetworkprob6} } \hspace{3pt}
    \subfloat[Network: Prior Conditional 2]{ \centering
    \includegraphics[width = 0.3\textwidth]{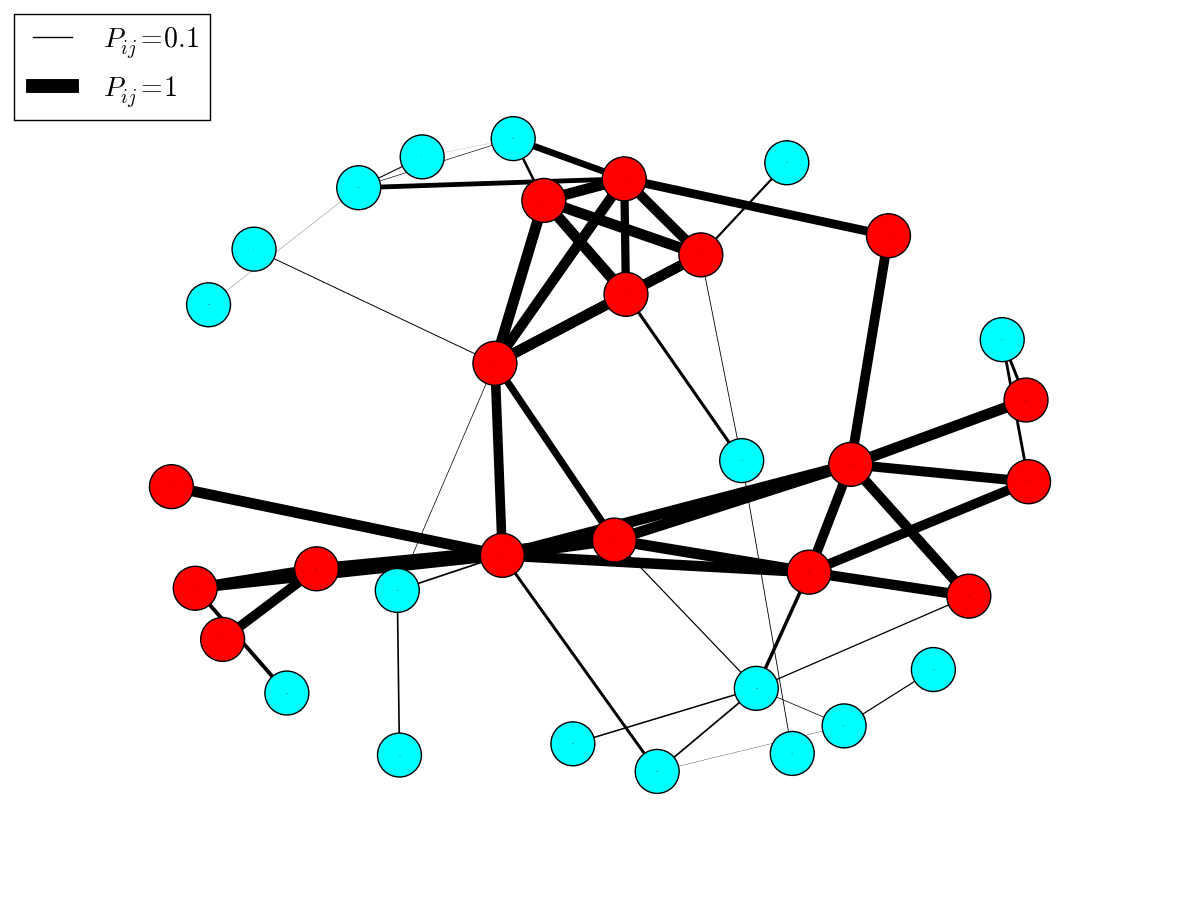}
    \label{TanzaniaNetworkprob12} }
  \caption{ \small{ The Tanzania network used to test the accuracy of constrained BL updates. Figure \ref{Tanzania_terrorists} shows a possible social network behind the terrorists responsible for the 1998 bombing of the US embassy in Tanzania. Figure \ref{TanzaniaNetworkprob6} and \ref{TanzaniaNetworkprob12} show the full networks used to test the BL model.  The thickness of the edge represents the probability of observing a relevant item on that edge, sampled from prior conditional 1 and prior conditional 2 for Figure \ref{TanzaniaNetworkprob6} and \ref{TanzaniaNetworkprob12} respectively.  The terrorist (irrelevant participants) are represented by red (blue) nodes and have true relevance 1 (0). }}
  \label{Tanzania_networks}
\end{figure}

The network shown in Figure \ref{Tanzania_terrorists} is a possible terrorist network associated with the bombing of the US embassy in Tanzania in $1998$ \citep{nevo2011information}.  The network consists of $17$ terrorists involved in the plot and is generated based on information from the Carnigie Mellon Computational Analysis of Social and Organisational Systems Laboratory \citeyearpar{casos2009}.  In addition to the $17$ terrorists, we also consider $17$ irrelevant participants. Edges are added randomly between an irrelevant participant's node and other nodes in the networks.  The full network used to test the accuracy of the constrained BL updates is given in Figure \ref{TanzaniaNetworkprob6}.  The true relevance of a terrorist (irrelevant participant) is set to $1$ ($0$). The true probability of observing a relevant intelligence item is sampled from the prior conditional beta distribution, given the involved participants' true relevancies.  These are shown by the width of the edge in Figure \ref{TanzaniaNetworkprob6} and Figure \ref{TanzaniaNetworkprob12}, for prior conditional 1 and 2 respectively. \\

For $250$ sets of $300$ observations, the updated expectations and variances are sequentially calculated in both the binary MRF model and constrained BL model (henceforth simply called the BL model).  The observations are on randomly selected edges and the relevance of the item is sampled from the true probability of observing a relevant item on that edge.  We look at the distribution of differences (MRF - BL) in the values of interest, after a given number of observations, on all nodes or edges in the network and over all 250 sets of observations.  For example, the difference in the expected relevance of participant $i$, after a set of observations $\mathbf{Y}$ would be given by $E[Z_i|\mathbf{Y}] - \hat{E}[Z_i|\mathbf{Y}]$.  \\

The binary MRF model is defined using the clique factor in Table \ref{prior_factor} with $[\lambda_1,\lambda_2] = [0.5,0.5]$ and using both prior conditional 1 and prior conditional 2.  For both prior conditional distributions, the difference between the BL model and the MRF model is small for the expectation of $\mathbf{Z}|\mathbf{Y}$, with the majority having an absolute difference of less that $0.1$ after $300$ observations.  However, the symmetry of the prior conditional distributions are reflected in the shape of the distribution of differences for $\mathbf{Z}|\mathbf{Y}$.  Prior conditional 2, which is more symmetric also has more symmetric differences.  Using prior conditional 1, the BL model is more likely to underestimate the expectation than overestimate.  \\

The accuracy of BL updates for $\mathbf{P} \mid \mathbf{Y}$ is affected more by the prior conditional probability.  There are only very small differences in the binary MRF and BL model updates of $\mathbf{P} \mid \mathbf{Y}$, when using prior conditional 1, compared with prior conditional 2. Prior conditional 2 is more dependent on the involved participants so a small error in the constrained BL expectation of $\mathbf{Z}\mid \mathbf{Y}$ will have a larger effect on the constrained BL updates of $\mathbf{P}\mid \mathbf{Y}$.  Hence the accuracy of the BL approximation is at least partially dependent on the model choice for the conditional probability distribution.  \\

\begin{figure}[h]
 \centering
\subfloat[Prior conditional 1: Differences  ]{ \centering
    \includegraphics[width = \textwidth]{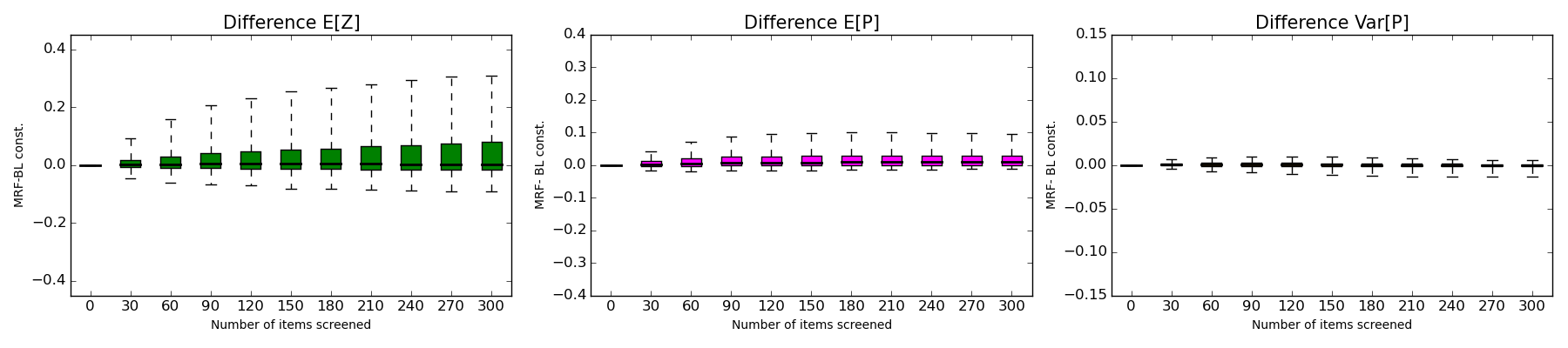}
       \label{Tanzania6_diffs} } \\
 \subfloat[Prior conditional 2: Differences ]{ \centering
    \includegraphics[width = \textwidth]{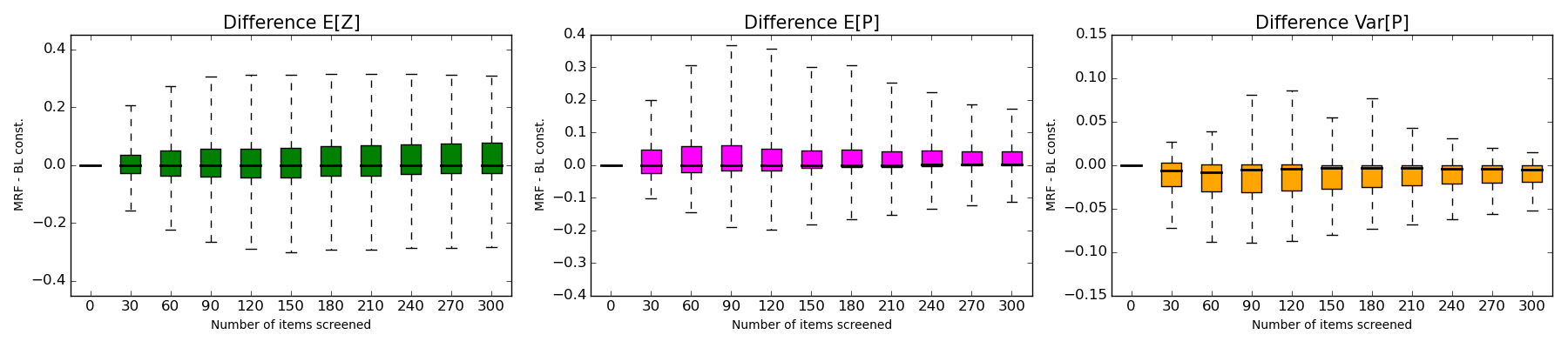}
    \label{Tanzania12_diffs} }
\caption{ \small{ The box plots show, for a given number of observations along the x-axis, the distribution of differences between the expectations and variances in the binary MRF model and using the constrained BL model.  Figure \ref{Tanzania6_diffs} show the results for prior conditional 1 and Figure \ref{Tanzania12_diffs} show the results for prior conditional 2 with $(\lambda_1,\lambda_2) = (0.5,0.5)$.   Each box plots show the median difference, and interquartile range.  The whiskers show the $95\%$ quantiles interval for the differences.  For both prior conditional probabilities, the majority of absolute differences are less than $0.1$.  } }
  \label{Tanzania_EZEPdiffs}
\end{figure}

\subsubsection{Sequential Decision Problem} \label{SecTanzaniaDecision}
For the BL model to be an appropriate approximate inference method, we would like the number of relevant items observed to be unaffected by which model is used to define the network.  For the networks used in Section \ref{SecTanzaniaDecision}, we simulate the number of items available on each edge from a Poisson($30$) distribution, and the number of relevant items from a binomial distribution using the true probability of being relevant on each edge as the probability parameter. Three heuristic algorithms are used in the search process; greedy, $\epsilon$-greedy and Bayes-UCB \citep{kaufmann2012bayesian}.  The greedy policy is a pure exploitation method, that will choose the edge with the highest expected probability of observing a relevant item on.  The $\epsilon$-greedy policy, will make a greedy selection with probability $1-\epsilon$, and with probability $\epsilon$ select a random edge; hence incorporating some exploration.  The greedy and $\epsilon$-greedy heuristic policies do not consider 
uncertainty in the estimate of the expected probability of observing a relevant item on each edge.  \\

Bayes-UCB policy has strong similarities with the upper confidence bounds used in UCB and its variants \citep{auer2002finite, garivier2011kl}. 
The algorithm uses upper quantiles of the posterior distribution of the expected reward on each action, and chooses the item which has the largest value of the appropriate quantile. We implement this method
by approximating the posterior distribution by a Gaussian distribution with the BL estimate of the posterior mean and variance. This method uses information on both the posterior mean and variance. Larger
variances will increase the value of the quantile used in the Bayes-UCB algorithm, and thus it encourages exploration of edges that have large uncertainty.  The Bayes-UCB algorithm used for the search process is given in Algorithm \ref{BayesUCB} in Appendix \ref{BayesUCB_Alg}.  \\

\begin{figure}[h]
 \centering
\subfloat[Prior Conditional 1 ]{ \centering
    \includegraphics[width=0.35\textwidth]{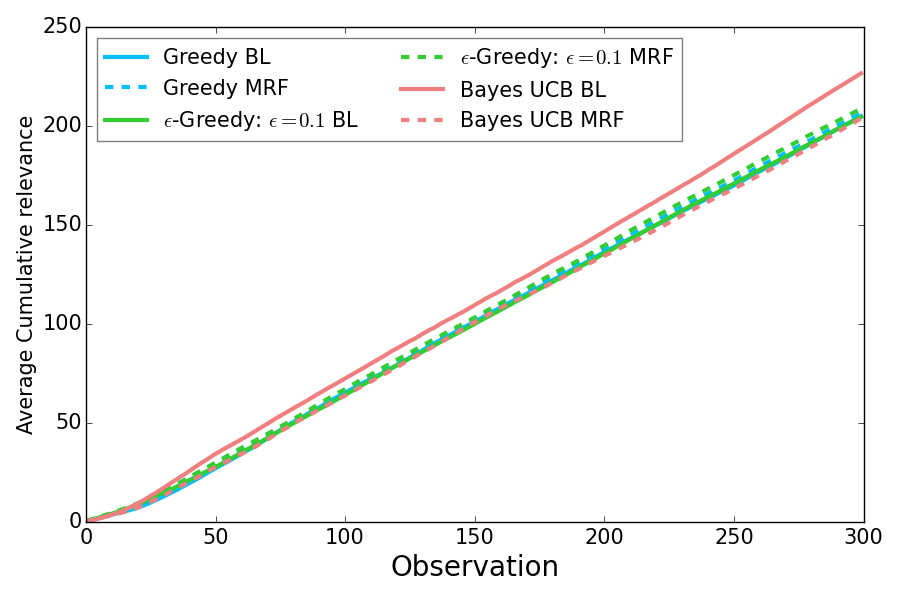}
    \label{Tanzania6_Dec_cumrel} } \hspace{5pt}
 \subfloat[ Prior Conditional 2]{ \centering
    \includegraphics[width=0.35 \textwidth]{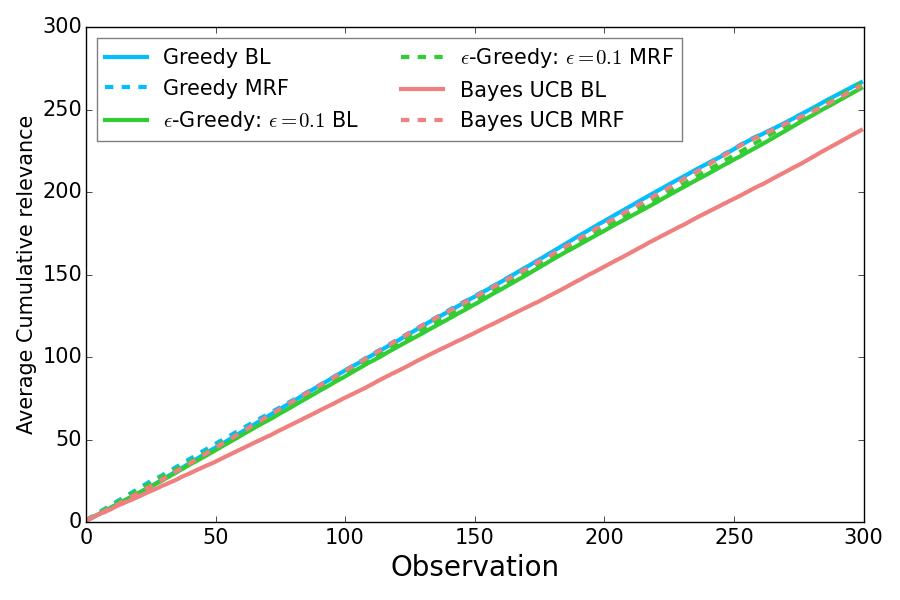}
    \label{Tanzania12_Dec_cumrel} }
  \caption{ \small{
The average cumulative number of relevant items found with different heuristic methods, using the BL model and binary MRF model.   The dotted lines are the cumulative number of relevant items observed using binary MRF model and the solid line is when the BL model is used.  Figure \ref{Tanzania6_Dec_cumrel} shows the cumulative average number of relevant items observed over $50$ runs of the decision problem, on the network with prior conditional 1 and Figure \ref{Tanzania12_Dec_cumrel} shows average cumulative average for the network using prior conditional 2.  There is very little difference between the performance of the heuristic methods for the BL model and binary MRF model.  }}
  \label{Tanzania_Dec_probs}
\end{figure}

Each heuristic method for the search process is run of $50$ times on the networks shown in Figure \ref{TanzaniaNetworkprob6} and \ref{TanzaniaNetworkprob12}. Figure \ref{Tanzania12_Dec_cumrel} shows the average cumulative number of relevant items observed over the $50$ repetitions using prior conditional 1, for the different heuristic methods using both the binary MRF model and BL model.  For each heuristic method, using the BL model as opposed to the binary MRF model results in roughly the same average number of relevant items observed.   Whilst the differences between expectations in the binary MRF model and BL model for prior conditional 2 were larger for $\mathbf{P}\mid \mathbf{Y}$, these errors have little effect on the performance of the heuristic methods, see Figure \ref{Tanzania12_Dec_cumrel}. This suggests the constrained BL updates may capture enough of the updating process to perform well in the heuristic methods. \\

\subsection{Correlated Nodes} \label{correlated_nodes}
Both the binary MRF model and BL model assume there will be correlation within the network.  Observing a relevant item on an edge will increase the probability of observing a relevant item on neighbouring edges.  An alternative model is to assume there is no correlation; each edge is independent of other edges in the network.  The probability of a relevant observation on each edge is given by a mixture beta distribution, conditioned on the involved participants' relevance values.  The prior probability for a pair of participants' relevance values is proportional to the clique factor in Figure \ref{prior_factor}. Henceforth, this model will be referred  to as the \textit{independent model}.  \\

For networks where there is some type of positive correlation between the relevance value of nodes in the network, we would expect the BL method to perform better than the independent model; the decision process will be able to make more informed decisions on where to search for future relevant items in the network based on observations on other edges. The method used to simulate the nodes relevancies is given in Section \ref{infect_simulation}.  A simple method is used to approximate the prior mean and covariance for the BL model, described in Section \ref{appox_cov}, and for the mean of this distribution, several prior clique factors are used for the independent model, see Section \ref{prior_indep}. Finally, a comparison of the update method's performances in the decision process for the simulated networks is given in Section \ref{corr_sim_results}.  

\subsubsection{Simulating Correlated Node Relevancies } \label{infect_simulation}
The method we use to simulate correlated random variables on the network, is to ``infect'' a node's neighbours with the same relevance value with some probability $\rho$.  A value of $\rho = 0.5$ will give a random allocation of node relevancies.  To simulate correlated random variables on the network:
\begin{enumerate}
 \item Pick an initial node, $Z_i$
 \item With probability $0.5$ let $Z_i = 1$ and $Z_i  = 0$ otherwise
 \item For each node $j \in ne(i)$, where $ne(i)$ are the neighbouring nodes of node $i$, which have not been assigned a relevance value, let
 \begin{align}
 Z_{j} = \left\{
  \begin{array}{ll}
    z_i & \text{with probability } \rho\\
    1-z_i& \text{with probability }  1 - \rho
  \end{array}
\right. . \label{fixedprobs}
\end{align}
 \item Randomly select an infected node with neighbours to infect, and go to 3).
\end{enumerate}

For the purpose of testing the effect of correlation in the network on the different models performance, given the node relevancies, we set the probability of observing a relevant item on an edge to:
\begin{align}
 P_{i,j} = \left\{
  \begin{array}{ll}
    0.0 & \text{if } z_i + z_j = 0\\
    0.2 & \text{if } z_i + z_j = 1\\
    0.9 & \text{if } z_i + z_j = 2
  \end{array}
\right.  .\label{fixedprobs}
\end{align}
A high value of $\rho$ will lead to a network which is more correlated.  One way to measure the spatial autocorrelation in the network is using Moran's I, which is given by:
 \begin{align}
  I = \frac{N}{\sum_{i,j} A_{ij}}\frac{ \sum_{i=1}^{N} \sum_{j=1}^{N} A_{ij} (y_i - \bar{y})(y_j - \bar{y})}{ \sum_{i=1}^{N} (y_{i}-\bar{y})^2},
 \end{align}
where $N$ is the number of random variables, $A$ is the adjacency matrix (or weight matrix), $y_i$ is the value of the $i$th random variable, and $\bar{y}$ is the mean of all random variables.  The expected value of Moran's I with no spatial autocorrelation is $\frac{-1}{N-1}$.  High autocorrelation will give a value close to $1$.

\subsubsection{Bayes Linear Prior Expectation and Covariance} \label{appox_cov}
The BL model requires prior specification of the mean and covariance of the $\mathbf{Z}$. In Section \ref{BLvsExact} these are calculated directly from the binary MRF model prior. When this model is not assumed to be the true underlying model, or on larger networks where calculating directly from the binary MRF is computationally prohibitive, we need to approximate the prior mean and covariance for the BL model.  The simplest way to approximate the prior mean is to assume all have the same value, $\mu$.\\ 

One method for approximating the prior covariance matrix is to consider the network structure \citep{loh2013structure}.  The precision matrix for random variables in a discrete MRF is graph structured.  For binary MRFs, non-zero entries of the precision matrix indicate an edge between the random variables in the associated graph. A zero entry in the precision matrix implies there is no edge between the corresponding nodes.  Let $ne(i)$ be the neighbouring nodes of node $i$ so that:
\begin{align}
 (\Sigma^{-1})_{ij} = 0 \text{ if} j \notin ne(i).
\end{align}
Using a factor, $\delta \in (-1,1)$, which controls the strength of correlation between random variables, a simple approximation to the covariance is given by:
\begin{align}
\Sigma = B^T Q^{-1} B, \label{approxcov_equ}
\end{align}
where:
\begin{align}
Q_{ij}  = \left\{
  \begin{array}{ll}
    1 & : i=j\\
    \frac{-\delta}{max(n_i,n_j)} & : i \in ne(j) \\
    0 & :  i \notin ne(j) 
  \end{array}
\right.,
\end{align}
and 
\begin{align}
 B = \begin{pmatrix}
      \sqrt{(\mu(1-\mu))/Q^{-1}_{1,1}} & 0 & \ldots & 0 \\
      0 & \sqrt{(\mu(1-\mu))/Q^{-1}_{2,2}} & \ldots & 0 \\
      \vdots & & \ddots & \vdots \\
      0 & \ldots & 0 & \sqrt{(\mu(1-\mu))/Q^{-1}_{m,m}}
     \end{pmatrix},
\end{align}
where $n_i$ are the number of neighbouring nodes for node $i$.  This method will ensure the diagonal entries of $Q$ are positive and the matrix is diagonally dominant; sufficient conditions for the matrix to be positive definite and hence ensuring a positive definite covariance matrix. It also ensures that the prior variance is consistent with the prior mean: $\Sigma_{ii} = \mu(1-\mu)$.

\subsubsection{Varying the Prior for the Independent Model} \label{prior_indep}

We need to also specify a prior for the independence model we compare BL to. For ease of comparison we try and match the priors of the two models, so they both have the same prior expectation.
However, for any value of prior expectation $E[Z] = \mu$, there is a continuous range of possible prior distributions for the independent model.  If we let $P_{ij}$ denote the prior probability of $Z_1=i$ and $Z_2=j$, 
then we impose the restrictions of symmetry, $P_{01} = P_{10}$, and that the mean matches the BL model mean: $P_{10} + P_{11} = \mu$.  This leaves one degree of freedom.  In order to ensure this prior distribution is not influencing the performance of the independent model when compared to the BL model, we can vary this final degree of freedom and run the independent model with a range of prior distributions.  We do this by varying the value of the 2nd moment $E[Z_iZ_j] = P_{11}$ between $0 \leq P_{11} \leq \mu$.

\subsubsection{Results on Simulated Networks} \label{corr_sim_results}
\begin{figure}[h]
 \centering
   \subfloat[$\rho = 0.5$]{\centering
      \includegraphics[width=0.18\textwidth]{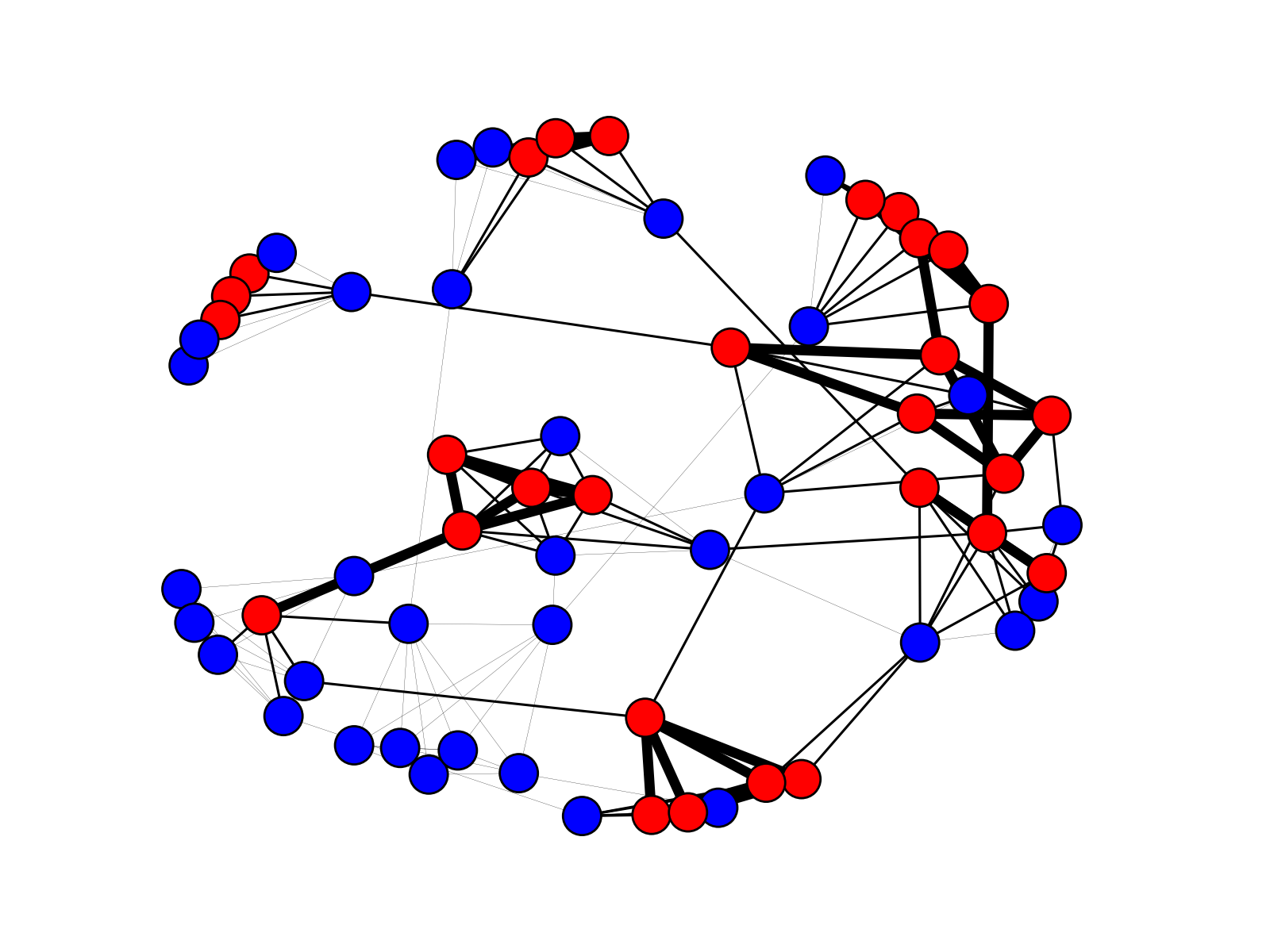}
 \label{cavemannet0}}  \hspace{1pt}
    \subfloat[$\rho = 0.6$]{\centering
      \includegraphics[width=0.18\textwidth]{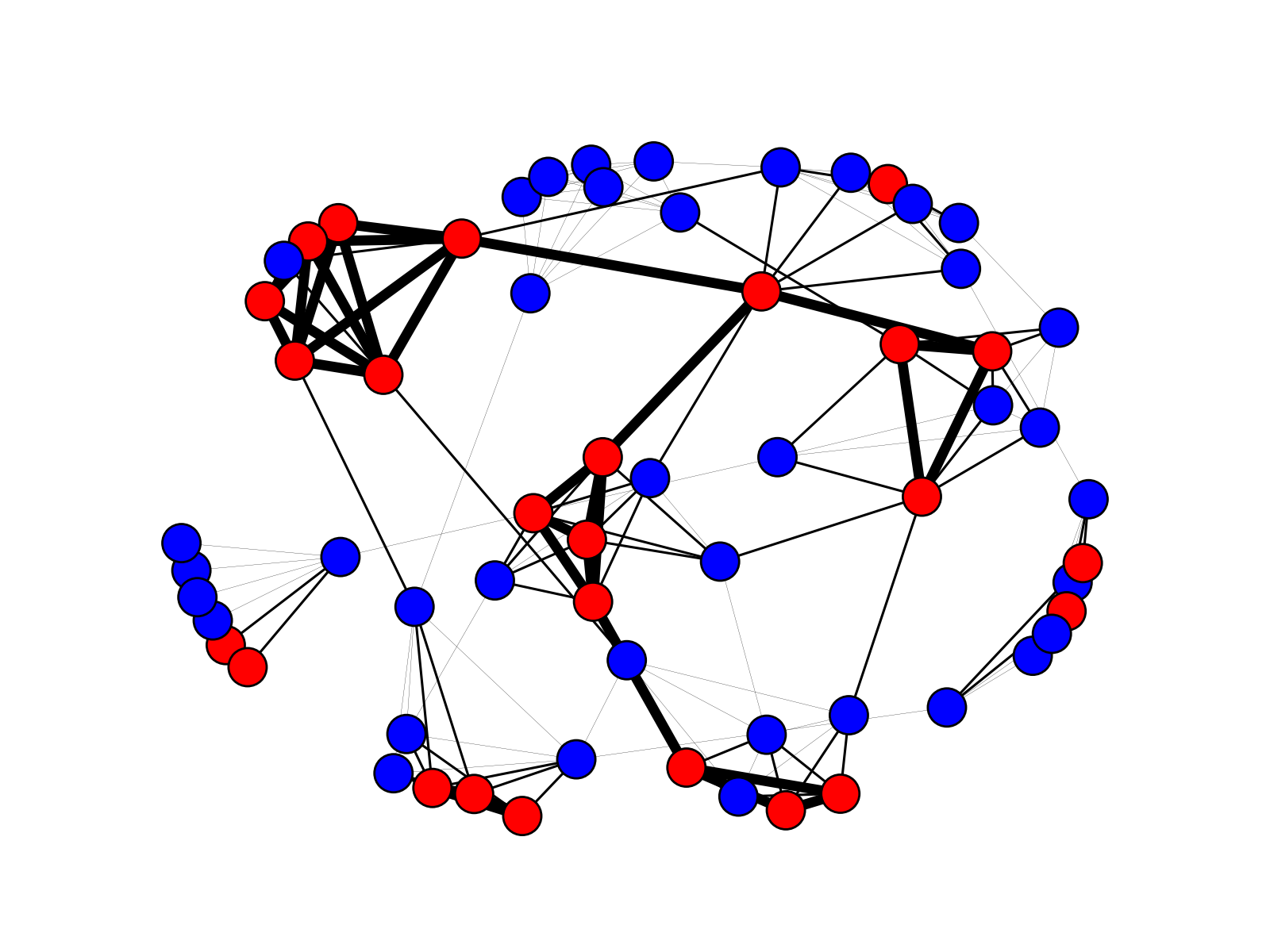}
 \label{cavemannet1}}  \hspace{1pt}
   \subfloat[$\rho = 0.7$]{\centering
      \includegraphics[width=0.18\textwidth]{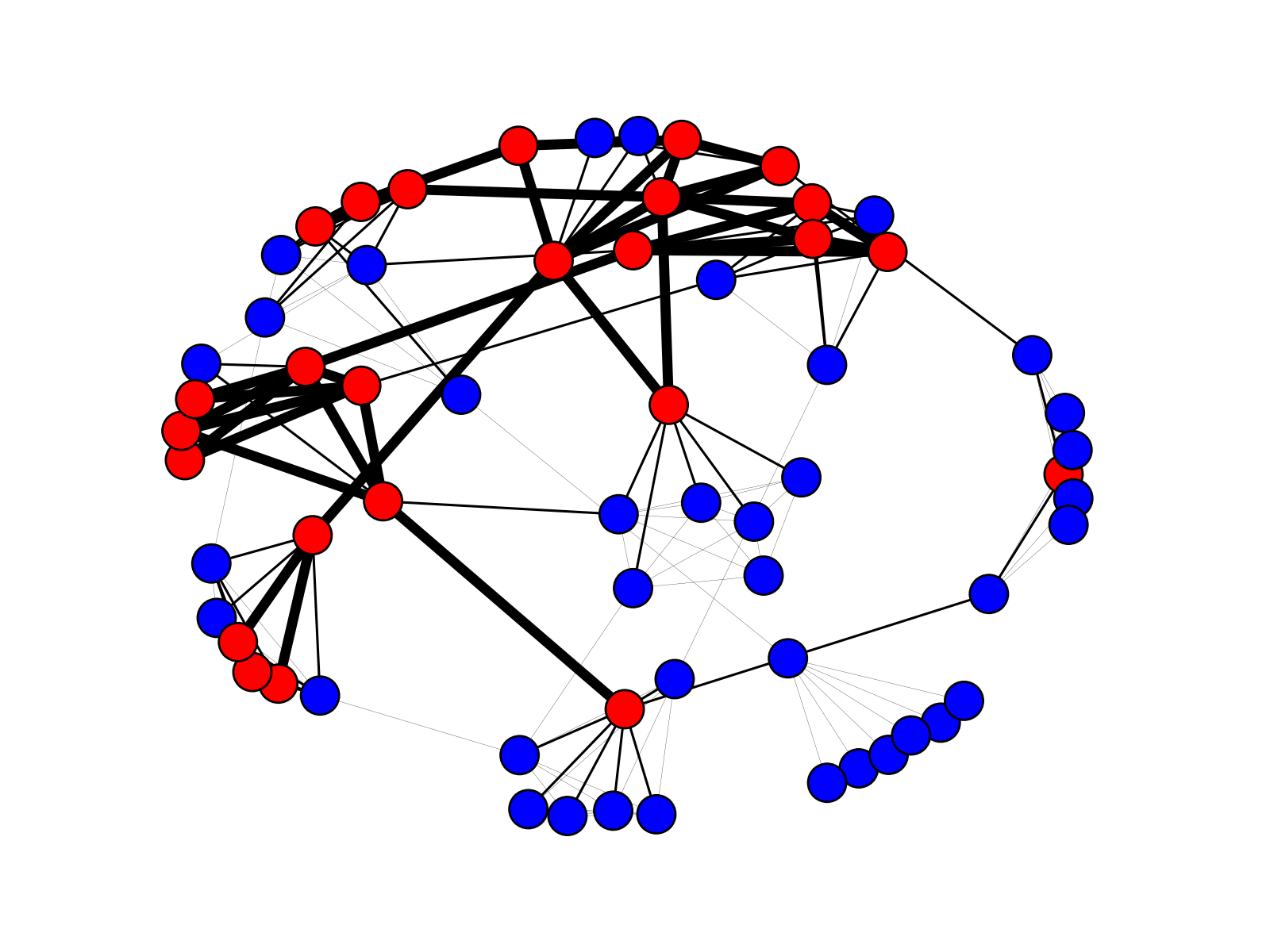}
 \label{cavemannet2}}  \hspace{1pt}
     \subfloat[$\rho = 0.8$]{\centering
      \includegraphics[width=0.18\textwidth]{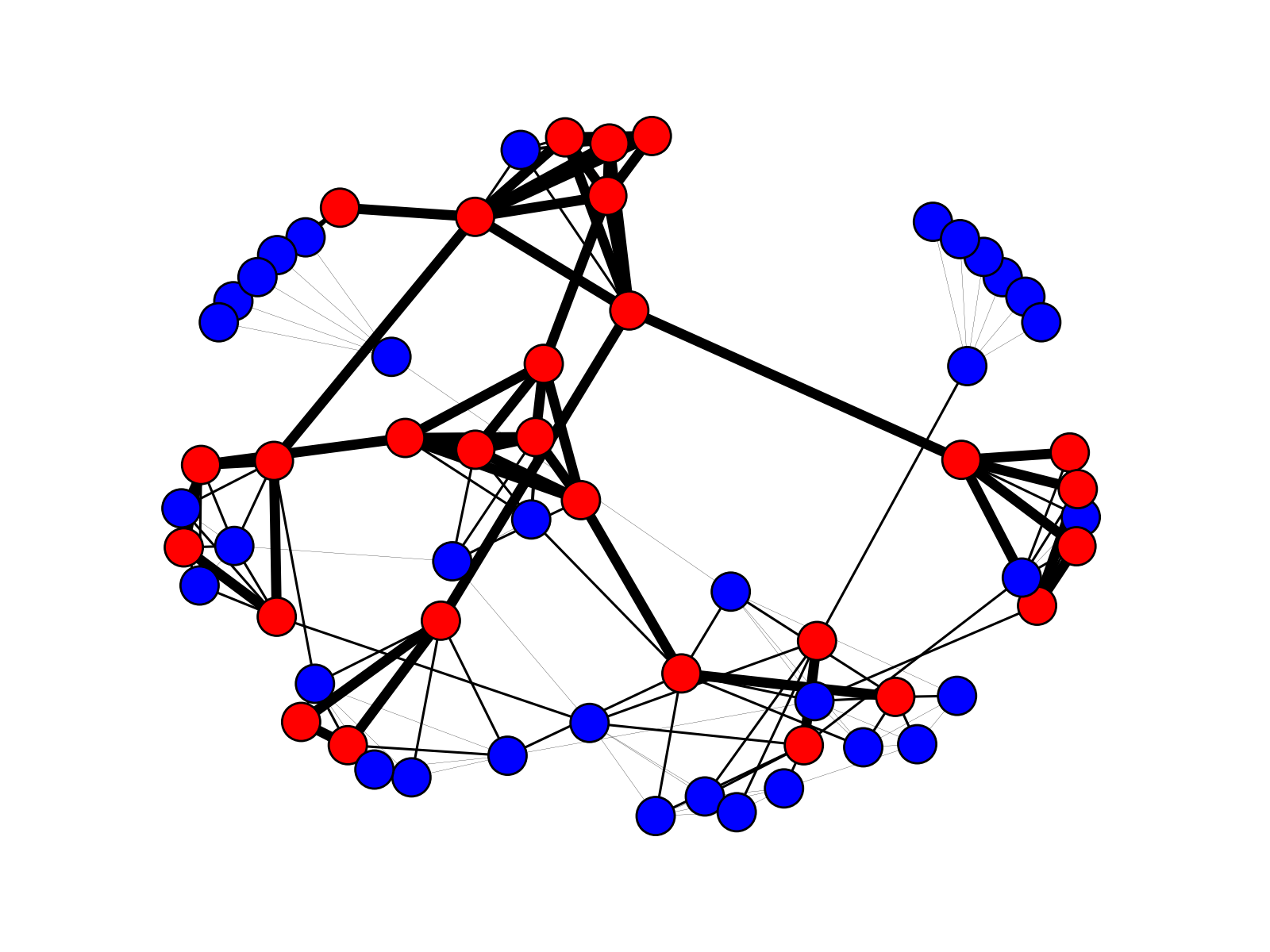}
 \label{cavemannet3}}  \hspace{1pt}
    \subfloat[$\rho = 0.9$]{\centering
      \includegraphics[width=0.18\textwidth]{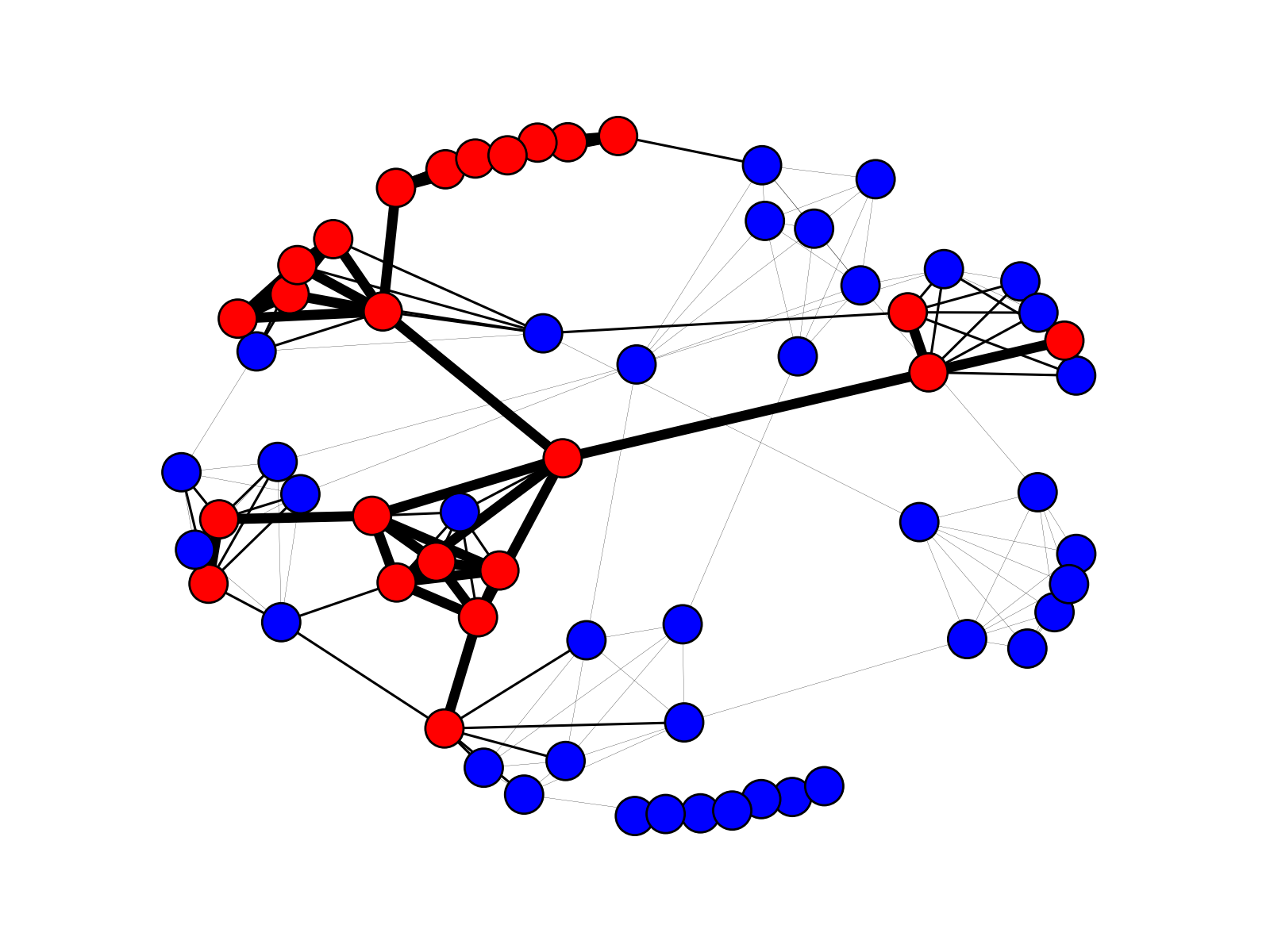}
 \label{cavemannet4}}  \hspace{1pt}
 \caption{\small{Relaxed caveman network and simulated relevance values of nodes, for different values of $\rho$ using the method in Section \ref{infect_simulation}. The red (blue) nodes represent relevant (irrelevant) participants.  For higher values of $\rho$, the nodes relevancies should be more clustered. } }
 \label{cavemannet}
\end{figure}

We look at the performance of the network based search method for a network simulated using the relaxed\_caveman\_graph function in NetworkX \citep{hagberg-2008-exploring}.  The true relevance of nodes is defined using the method in Section \ref{infect_simulation} for a range of $\rho$ values.  For a value of $\rho$, the same network is used in all iterations, see Figure \ref{cavemannet}.  Table \ref{Table_MoransI} gives the Moran's I values for both the nodes and edges in the networks.  As the value of $\rho$ increases, the Moran's I for the nodes, $I_{nodes}$, and edges, $I_{edges}$, both increase.  We would expect the BL model to have superior performance for networks where there is correlation in the edge random variables.  Hence we would expect the BL model to perform better on the networks with higher $I_{edges}$.\\

\begin{table}[h]
 \centering
 \begin{tabular}{|c|cc|} \hline
  $\rho$ &  $I_{edges}$ & $I_{nodes}$ \\ \hline
    $0.5$ & $0.44$ & $0.09$ \\
    $0.6$ & $0.52$ & $0.47$ \\
     $0.7$ & $0.61$ & $0.28$ \\
    $0.8$ & $0.62$ & $0.25$  \\
     $0.9$ & $0.76$ & $0.5$\\ \hline
 \end{tabular}
\caption{\small{Moran's I values for the nodes, $I_{nodes}$, and edges, $I_{edges}$, for the networks shown in Figure \ref{cavemannet}.  As the value of $\rho$ increases, the Moran's I values generally also increase.}}
\label{Table_MoransI}
\end{table}

\begin{figure}[h]
 \centering
   \subfloat[greedy: $\mu = 0.25$]{\centering
      \includegraphics[width=0.35\textwidth]{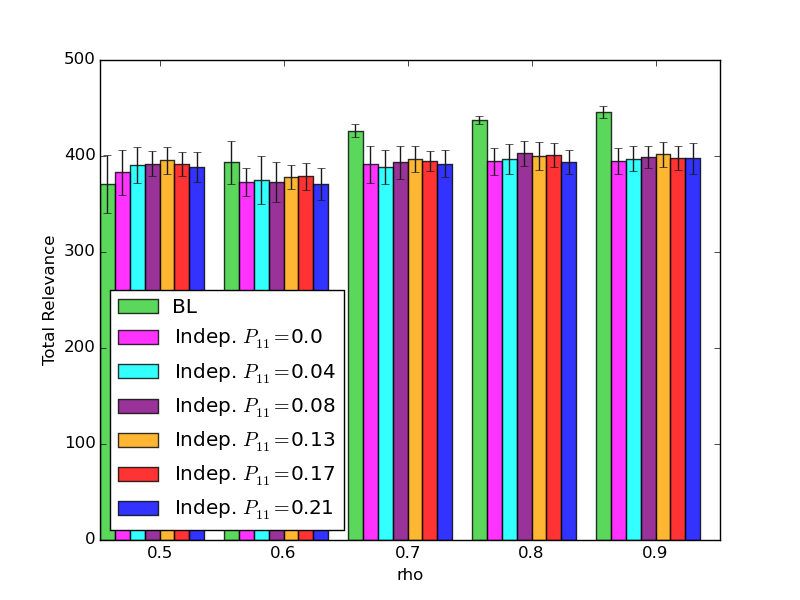}
 \label{NNS_range_greedy_caveman_500_EZ025}}  \hspace{5pt}
    \subfloat[greedy: $\mu = 0.5$]{\centering
    \includegraphics[width=0.35\textwidth]{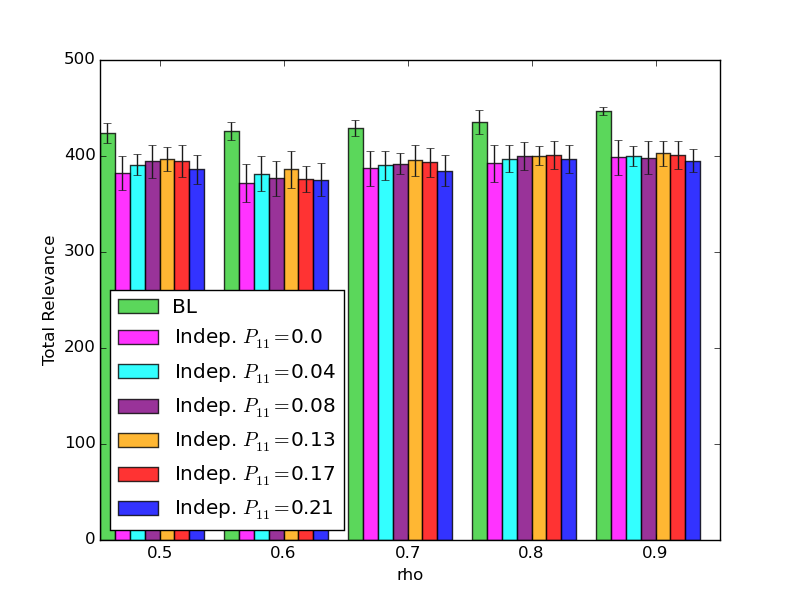}
 \label{NNS_range_greedy_caveman_500_EZ05}} 

 \caption{\small{The mean and $\pm 1$ standard error bars of the total number of relevant items found over $500$ observations for the greedy policy run on networks simulated using the relaxed\_caveman\_graph function in NetworkX \citep{hagberg-2008-exploring}, shown in Figure \ref{cavemannet}.  The independent models are run for a range of second moments with the mean matching the BL model mean.  The BL model performs better than the independent model for most values of $\rho$. } }
 \label{NNS_range_caveman_greedy}
\end{figure}

For both the independent model and the BL model, we set the prior conditional probability distribution to that of Table \ref{prior_prob2}.  The prior mean for the BL model is set to either $0.25$ or $0.5$ and the covariance is defined using the method in Section \ref{appox_cov} with $\delta = 0.8$.  The independent model is run for a range of prior distributions as described in Section \ref{prior_indep}, matching the mean of the BL model.  Figure \ref{NNS_range_caveman_greedy} shows the mean total number of relevant observations for BL model and the independent model using the greedy decision policies over 50 repetitions each with 500 observations. For $\rho = 0.7$, $\rho = 0.8$ and $\rho = 0.9$, the simulated networks have high Moran's I values for both the nodes and the edges.  This is reflected in the performance of the BL model. However, even when there is little correlation in the networks, the BL model tends not to be detrimental.  A very similar number of relevant items are screened for 
the independent model for all prior distributions considered.

\subsection{Enron network} \label{SecEnronDecision}
In this section we look at how the different models perform, when applied to communication networks simulated from the Enron Corpus.  The Enron Corpus consists of hundreds of thousands of emails from over 150 Enron employees at the Enron Corporation. The dataset was made public during the US government’s legal investigation of Enron after it's collapse in 2001 and provides a good example of the type of data set used with legal cases.  This data set has similarities with the datasets in corporate legal cases such as \textit{Oracle America, Inc. v. Google Inc.} case. It was used as evidence in as legal investigation as evidence of knowledge of events which were taking place within the companies.   \\

The communication networks are simulated from the Enron Corpus using from \cite{ellis2013algorithms}.  An email in the network is classed as relevant if it contains the works ``Money'' or ``Finance''.  For a relatively small network, we compare the total number of relevant items observed using the binary MRF model, the BL model and the independent model, see Section \ref{Enron300}.  For larger networks, where the binary MRF model is computationally intractable, we look at the performance of the BL model and the independent model, see Section \ref{enronlarge}.  For the decision algorithms, we consider the 
greedy and $\epsilon$-greedy, with $\epsilon = 0.1$, policies.  \cite{nevo2011information} suggests these algorithms perform surprisingly well, and that exploitation for sequential decision problems on networks is less important.  We also consider the Bayes-UCB policy, which uses the variance of the rewards as well as the expectation

\subsubsection{Small Enron Network} \label{Enron300}
\begin{figure}[h]
 \centering
    \subfloat[Enron Network]{\centering
    \includegraphics[width=0.35\textwidth]{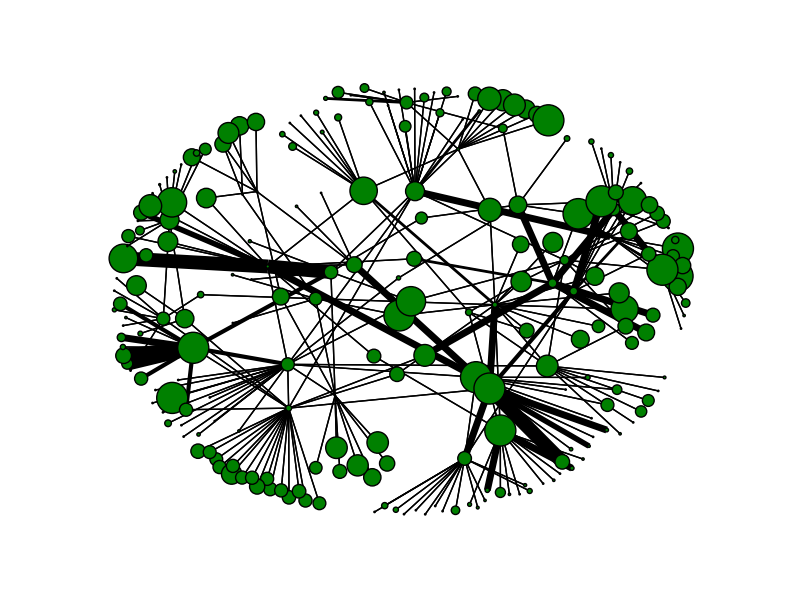}
  \label{EnronNetwork}} \hspace{3pt}
  \subfloat[Bargraph of total relevance ]{\centering
    \includegraphics[width=0.35\textwidth]{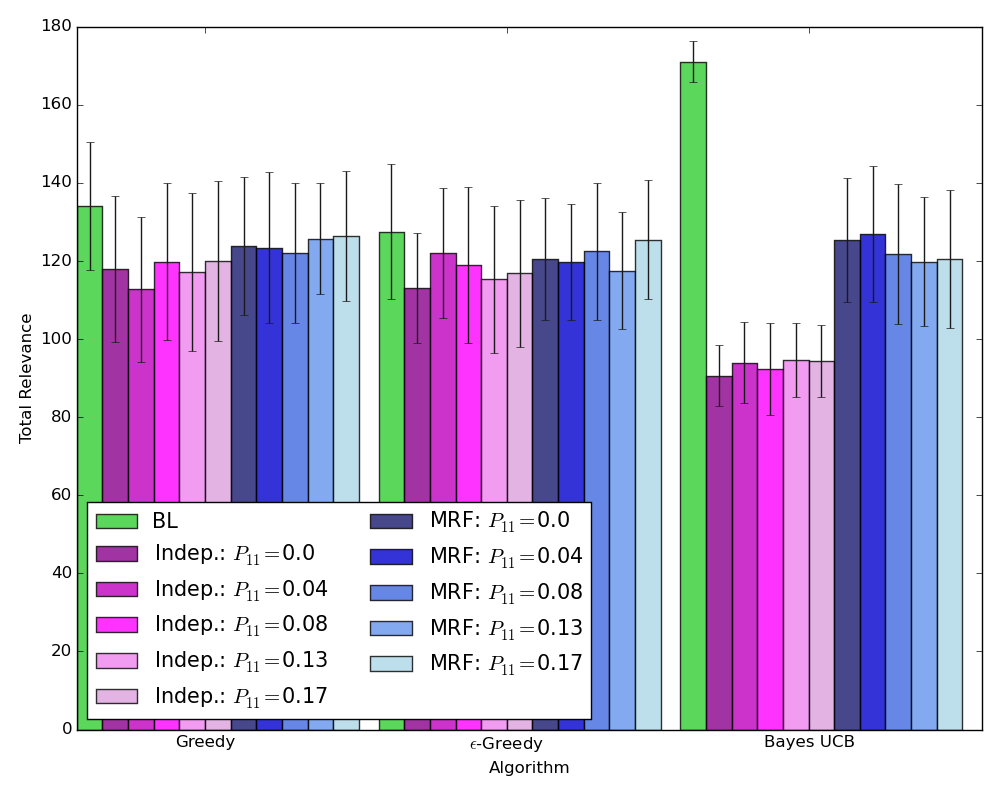}
  \label{Lisa300_bargraph}} \\
  \subfloat[Average Cumulative Relevance: greedy policy ]{\centering
    \includegraphics[width=0.35\textwidth]{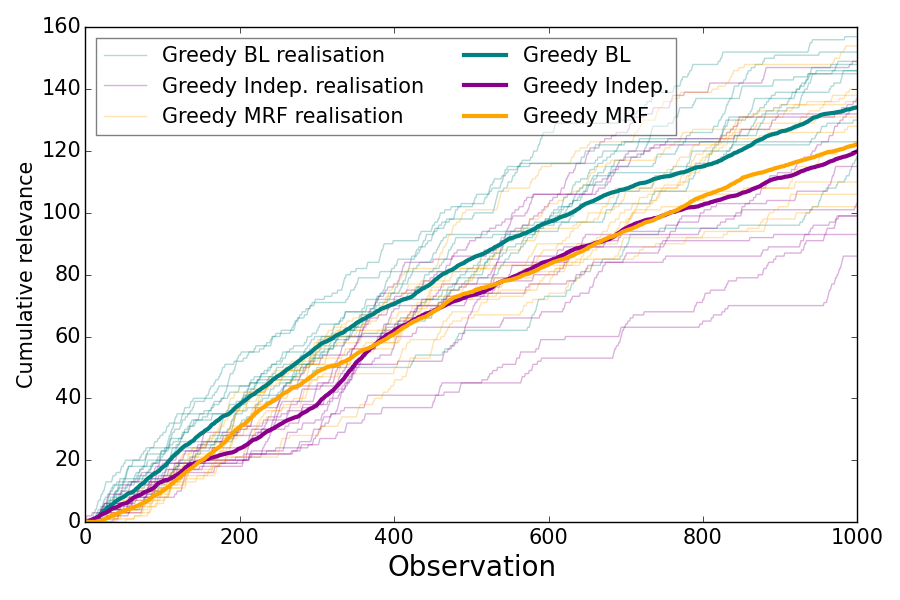}
  \label{Lisa300_cumulativerel}} \hspace{3pt}
  \subfloat[Number of Edge Changes: greedy policy ]{\centering
    \includegraphics[width=0.35\textwidth]{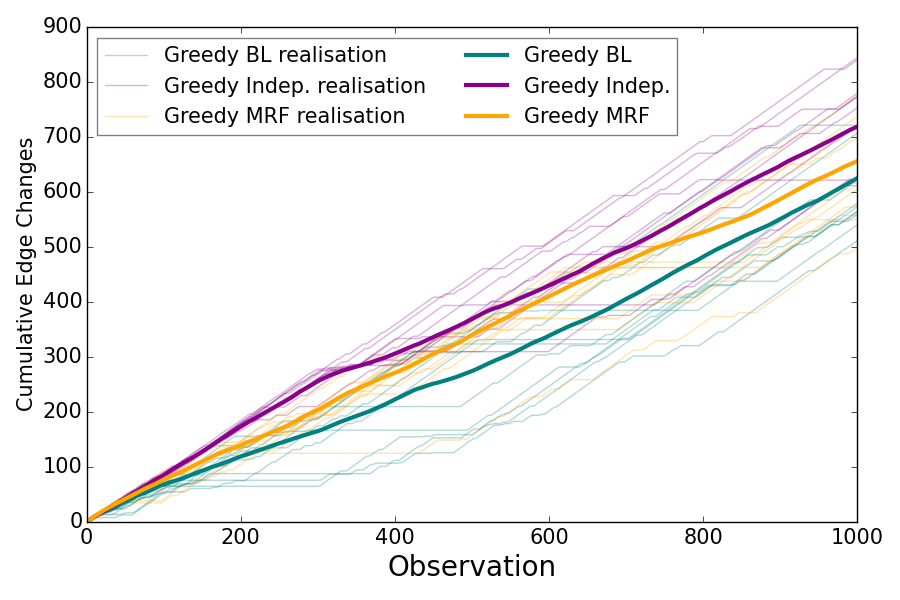}
  \label{Lisa300_cumulativechange}} \hspace{3pt}
  \caption{\small{Results for a small Enron network, shown in Figure \ref{EnronNetwork}. Figure \ref{Lisa300_bargraph} shows the mean and $\pm 1$ standard error bars of the total number of relevant items found over $1000$ observations for different models and heuristic policies.  The mean cumulative relevance over $1000$ observations is shown in Figure \ref{Lisa300_cumulativerel} for the greedy policy with $P_{11} = 0.13$. Figure \ref{Lisa300_cumulativechange} shows the mean number of times the policy changes edges. The BL model performs better to both the MRF and independent model for each of the decision policies.}}
  \label{EnronLisa300_results}
\end{figure}

For a relatively small Enron network, we can compare the BL model with both the binary MRF model and the independent model.  The network considered has $234$ nodes and $275$ edges and is shown in Figure \ref{EnronNetwork}.  There are a total of $4958$ email in the network, of which only $260$ are relevant to the intelligence query. For the prior BL model we set $E[Z] = 0.25$ and use an approximate covariance with $\delta = 0.8$.  A range of priors for the binary MRF model and the independent model are run by matching the mean of the BL model and varying the second moment of the clique factor, see Section \ref{prior_indep}. \\

We run $30$ repetitions with $1000$ observations on each repetition.  More relevant items are observed using the BL model than the independent model or the MRF model.  The Bayes-UCB policy with BL updates gives far superior performance to the other two methods.  The independent model does particularly badly with this policy.  For the prior clique factor with $P_{11} = 0.13$ and using the greedy policy, Figure \ref{Lisa300_cumulativerel} and \ref{Lisa300_cumulativechange} shows the mean number of relevant observations and the number of times a change of edge is made in the algorithm.  The BL method tends to stick with edges for longer than the independent model or the MRF model.  

\subsubsection{Larger Enron Networks} \label{enronlarge}
\begin{figure}[h]
 \centering
    \subfloat[$448$ nodes and $630$ edges ]{\centering
     \includegraphics[width=0.3\textwidth]{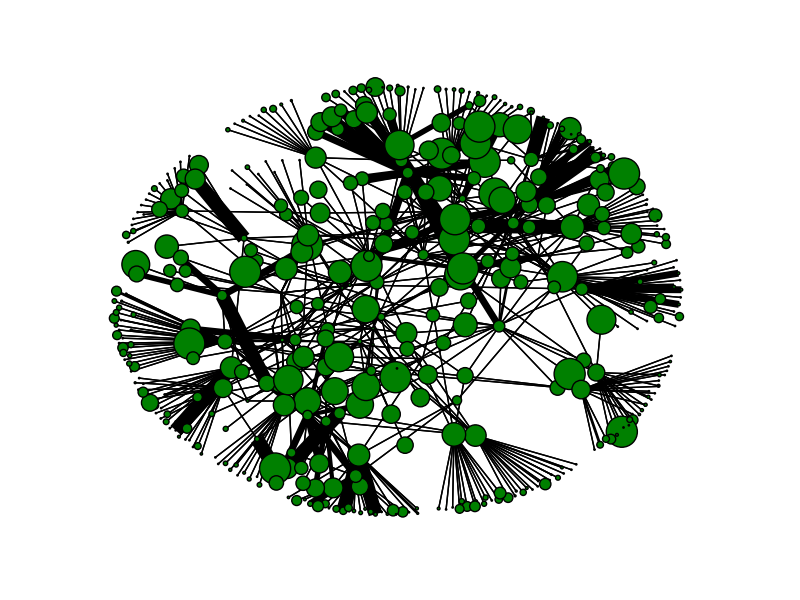}
   \label{EnronNetwork_500_2}} \hspace{3pt}
   \subfloat[$669$ nodes and $1133$ edges]{\centering
    \includegraphics[width=0.3\textwidth]{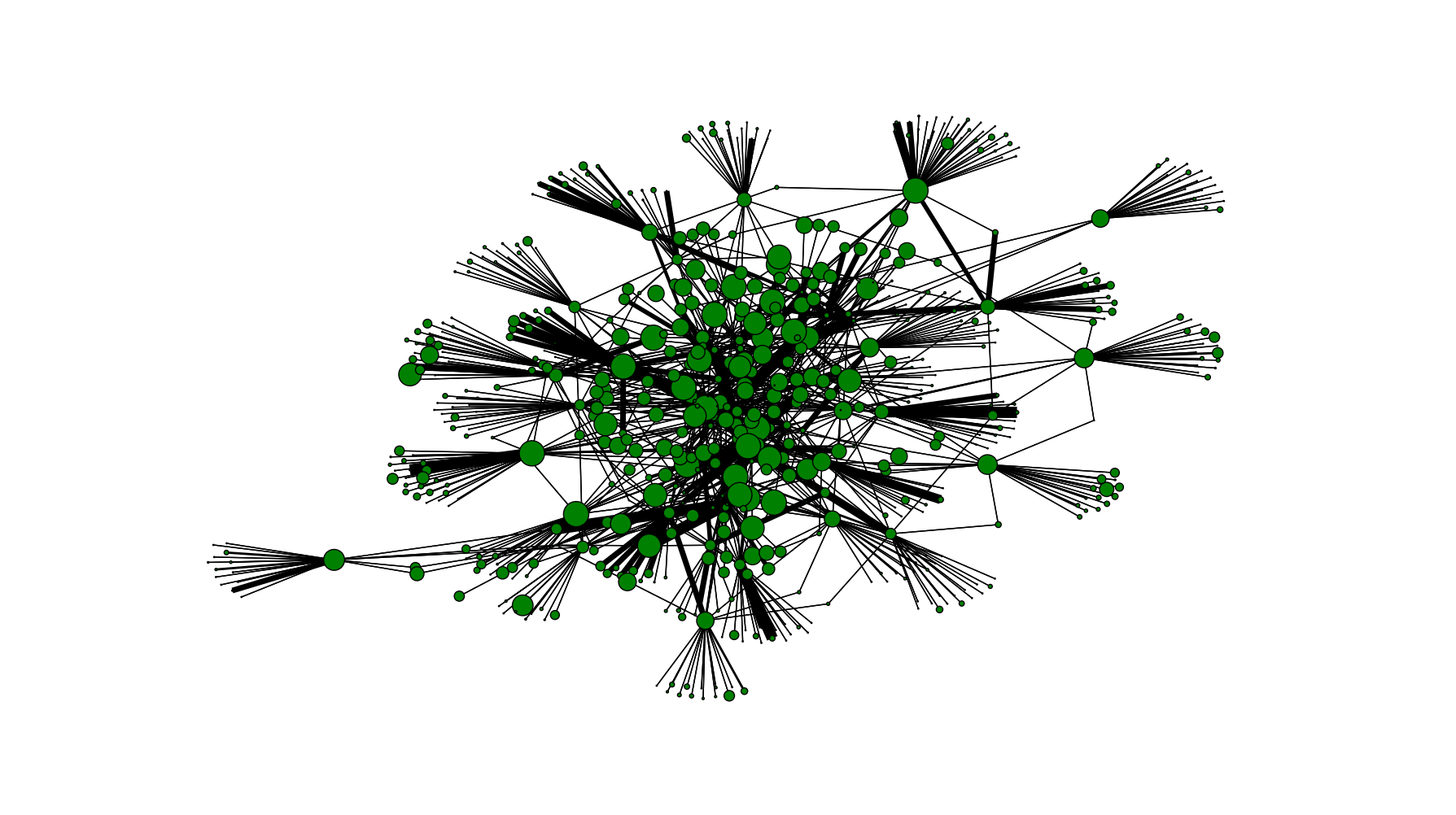}
  \label{EnronNetwork_700}} \hspace{3pt}
   \subfloat[$1641$ nodes and $1957$ edges]{\centering
    \includegraphics[width=0.3\textwidth]{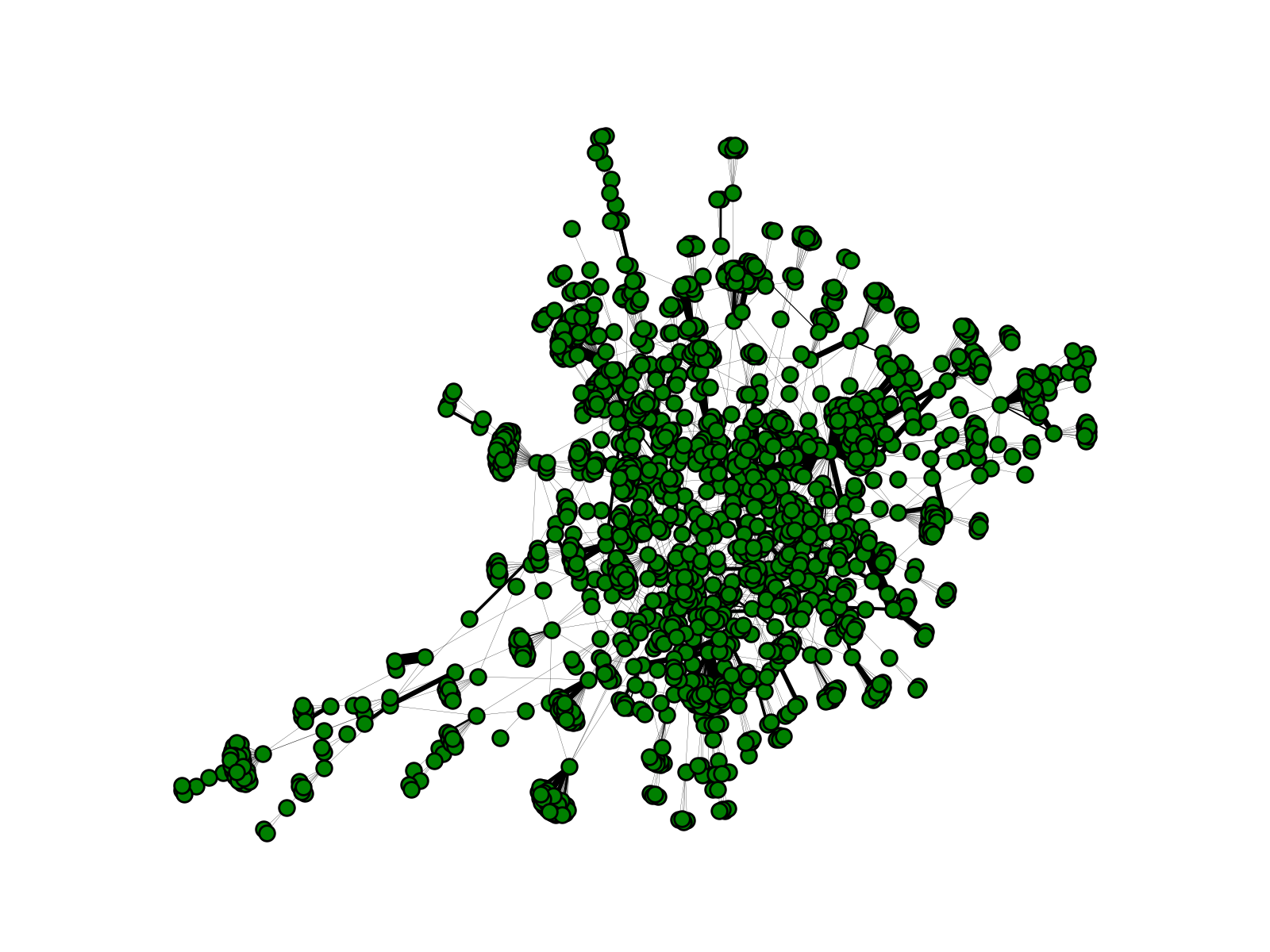}
  \label{EnronNetwork_2000}}\\
 \subfloat[Cumulative Relevance: $448$ nodes]{\centering
    \includegraphics[width=0.30\textwidth]{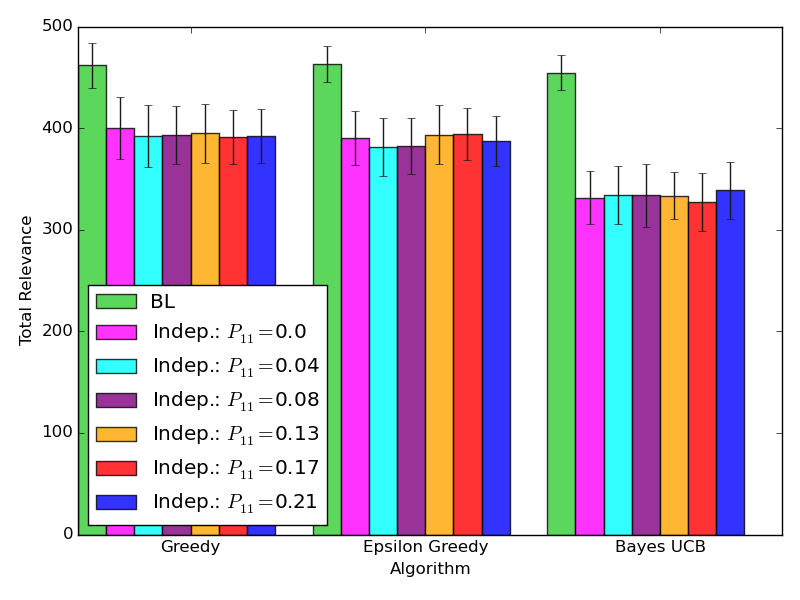}
  \label{NNS_range_500_2}} \hspace{3pt}
 \subfloat[Cumulative Relevance: $669$ nodes]{\centering
    \includegraphics[width=0.30\textwidth]{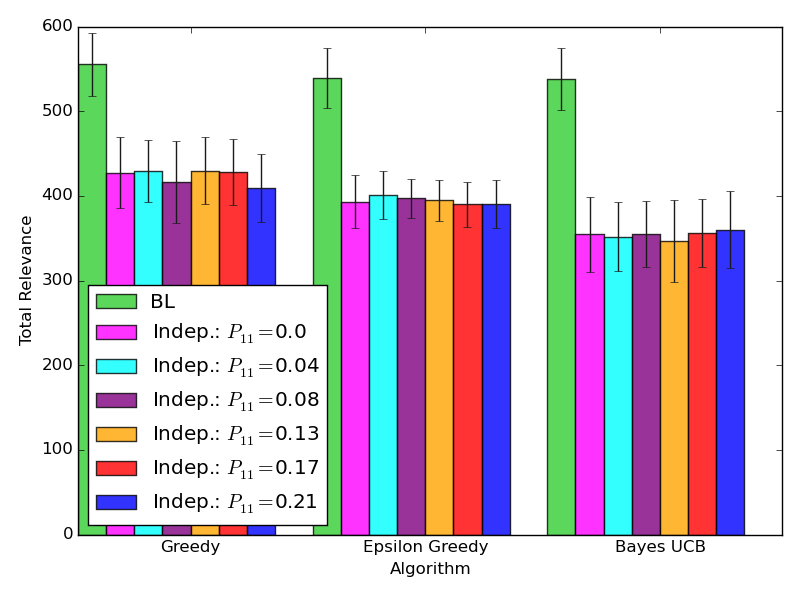}
 \label{NNS_range_700}} \hspace{3pt}
  \subfloat[Cumulative Relevance:$1641$ nodes]{\centering
    \includegraphics[width=0.30\textwidth]{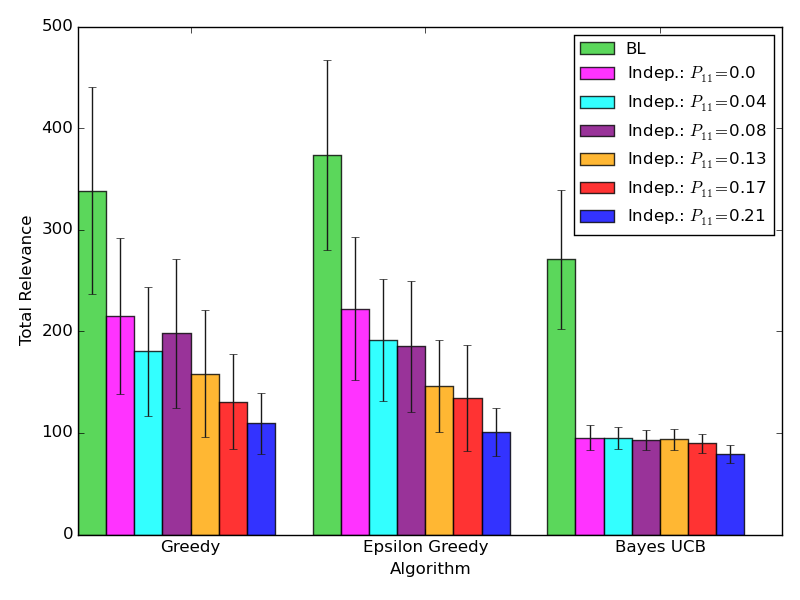}
 \label{NNS_range_2000}} 
 \caption{\small{The mean and $\pm 1$ standard error bars for the total number of relevant items found for the greedy, $\epsilon$-greedy and Bayes-UCB policies run on a subsets of the Enron Corpus for the BL model and independent models with a range of prior clique factors, chosen to match the mean of the BL model prior. For each network the BL model performs better than the independent model.} }
 \label{NNS_range_Enron}
\end{figure}

For larger networks, the binary MRF model is computationally intractable.  We run the BL model and independent model on three networks.  The network shown in Figure \ref{EnronNetwork_500_2} has $448$ nodes and $630$ edges.  Figure \ref{EnronNetwork_700} has $669$ nodes and $1133$ edges. Finally, Figure \ref{EnronNetwork_2000} has $1641$ nodes and $1957$ edges. For all networks, we set the prior mean to $E[Z] = 0.25$ in the BL model with the prior covariance approximated with $\delta = 0.8$, using the method in Section \ref{appox_cov}.  The independent model is run with a range of priors by fixing $E[Z_iZ_j] = P_{11}$ over the range of $0 < P_{11} < \mu $ where $\mu = 0.25$.  The range of $P_{11}$ values is given by $P_{11} = [0.0, 0.04, 0.08, 0.13, 0.17, 0.21]$. \\

For the two smaller networks we get very similar results for total number of relevant items observed for the independent model, with all values of $P_{11}$.  The BL model gives superior performance to the independent model. For the largest Enron network, the independent model's performance depends on the value of $P_{11}$.  The model where $P_{11} = 0.0$ performs best; that is when we assume there is zero probability that both nodes involved are relevant to the query.  The BL model performs better than the independent model for all prior clique factors. All three decision algorithms give similar results.  If we were to increase the number of items observed, the independent model eventually catches up with the BL model as it finds the edges with higher probabilities of observing relevant items. \\

\begin{figure}
\centering
\subfloat[Cumulative Relevance for greedy policy]{\centering
    \includegraphics[width=0.4\textwidth]{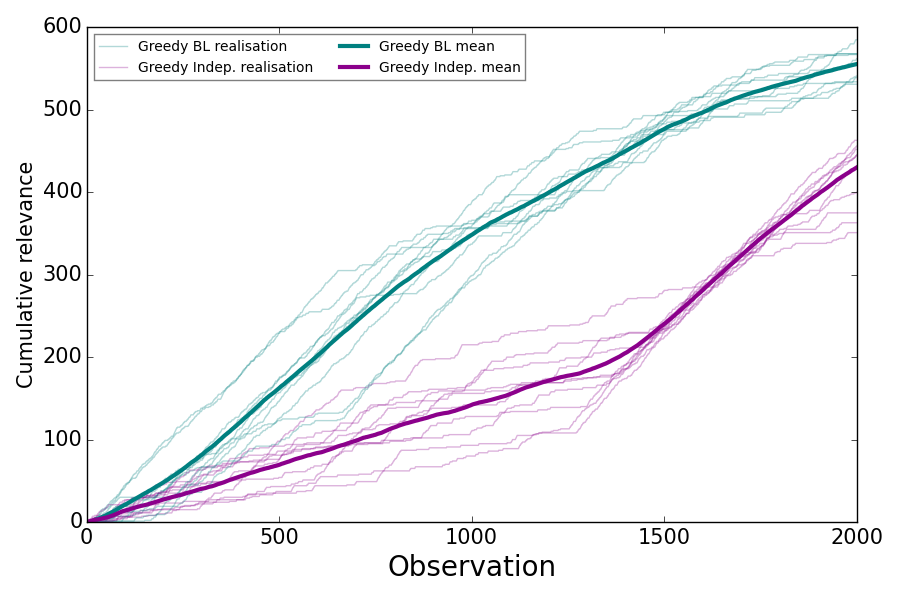}
  \label{Lisa700_cumulativerel_greedy}} \hspace{5pt} 
 \subfloat[Cumulative Number of Edge Changes for greedy policy]{\centering
    \includegraphics[width=0.4\textwidth]{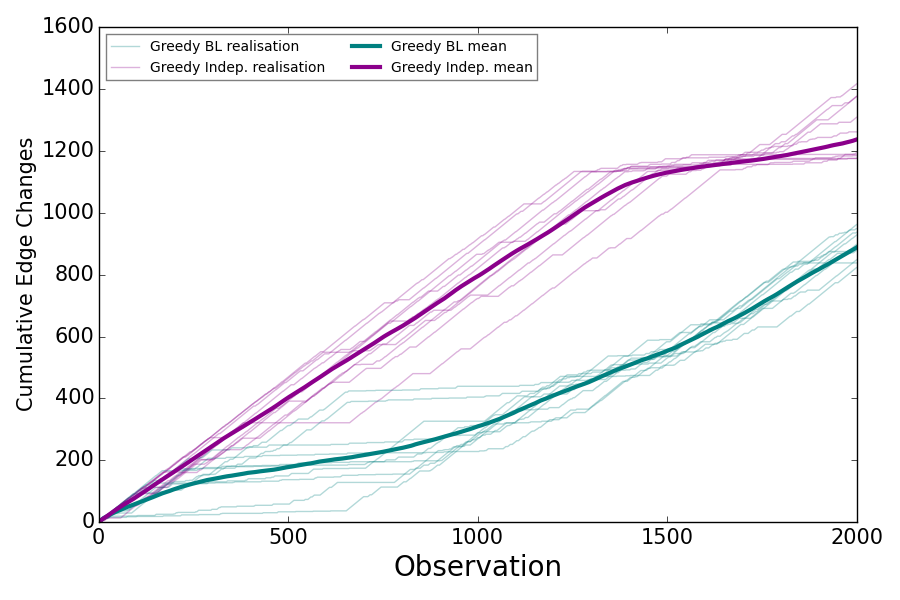}
  \label{Lisa700_cumulativechanges_greedy}} \\
   \subfloat[Distribution of times an item on an edge in the top 95th quantile is observed]{\centering
    \includegraphics[width=0.4\textwidth]{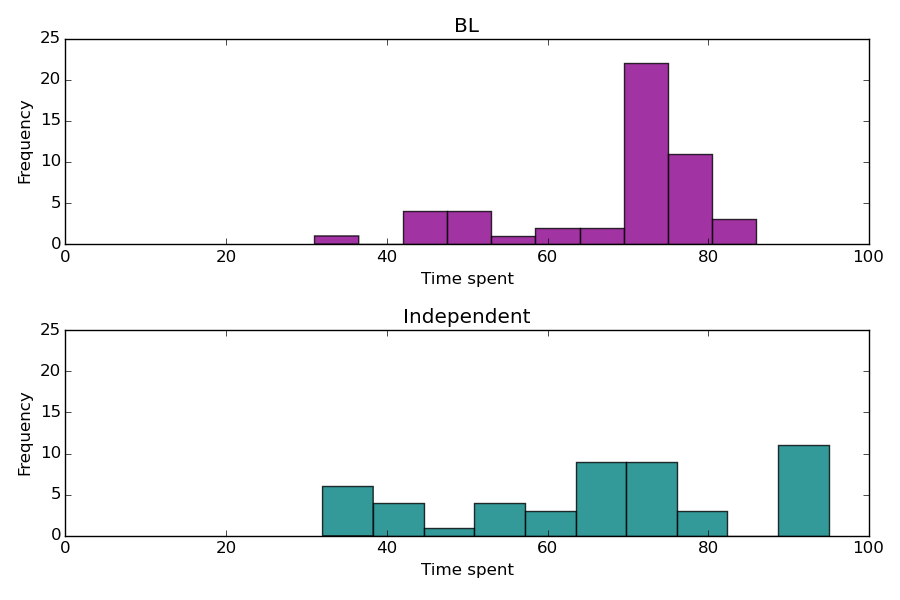}
  \label{Lisa700_histogram_goodedges95}} \hspace{5pt} 
  \subfloat[Distribution of times an item on an edge in the top 90th quantile is observed]{\centering
    \includegraphics[width=0.4\textwidth]{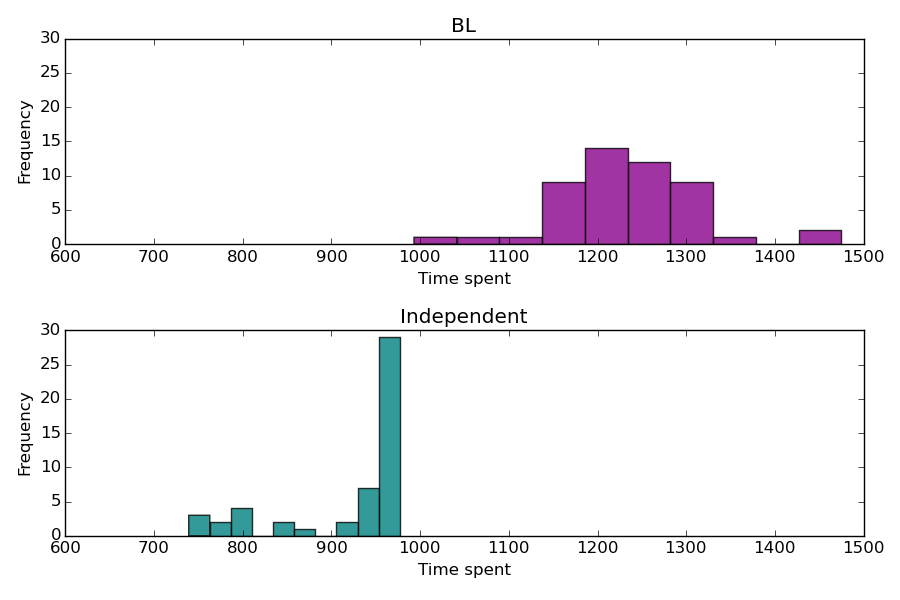}
  \label{Lisa700_histogram_goodedges90}} \\
     \subfloat[Cumulative number of observations an edge in the top 95th quantile]{\centering
    \includegraphics[width=0.4\textwidth]{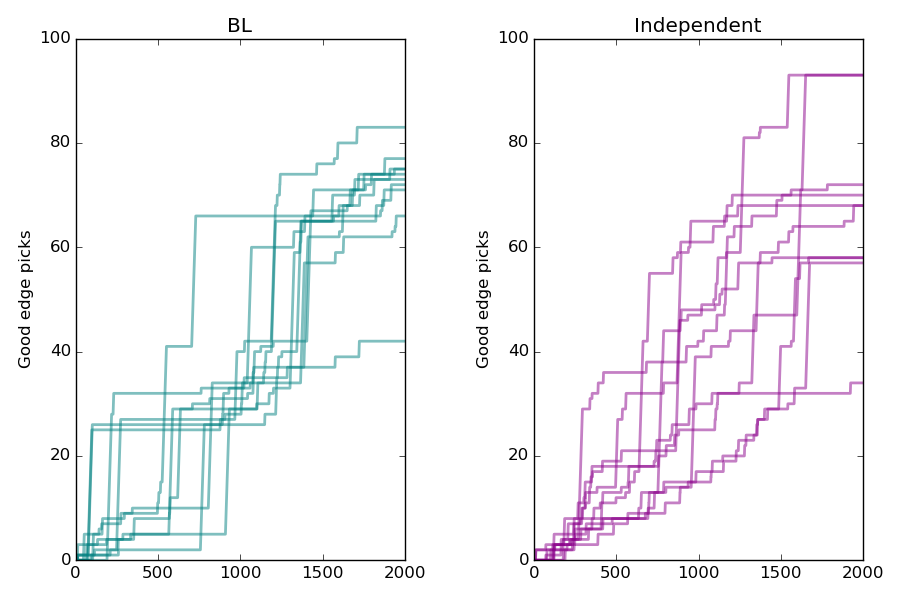}
  \label{Lisa700_realisations_goodedges95}} \hspace{5pt}
     \subfloat[Cumulative number of observations an edge in the top 90th quantile]{\centering
    \includegraphics[width=0.4\textwidth]{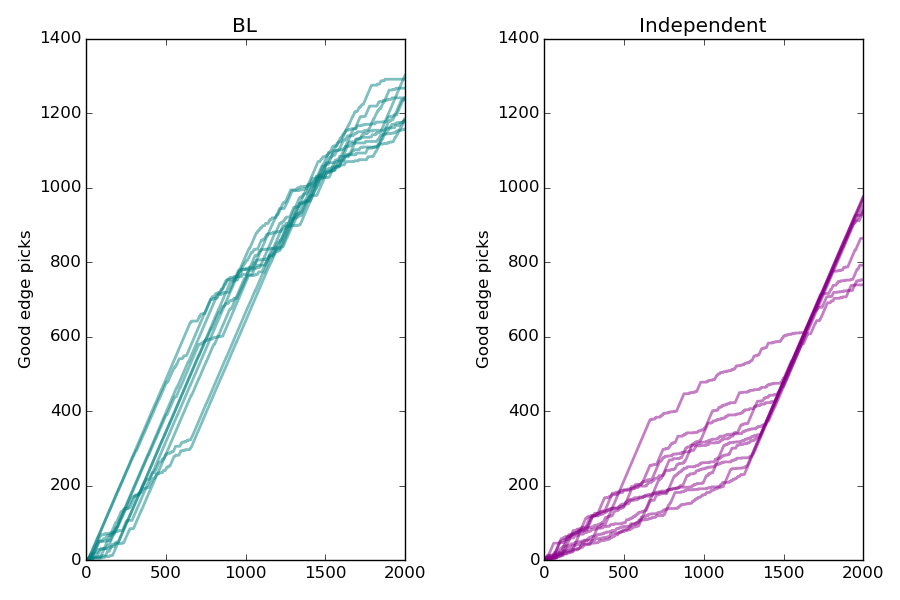}
  \label{Lisa700_realisations_goodedges90}} 
  \caption{\small{ Results for the greedy policy for the network shown in Figure \ref{EnronNetwork_700}, with $669$ nodes and focusing on the independent model with $P_{11} = 0.13$. Figure \ref{Lisa700_cumulativerel_greedy} shows the cumulative number of relevant items observed.  Figure \ref{Lisa700_cumulativechanges_greedy} shows the cumulative number of times the algorithm changed edges. Figure \ref{Lisa700_histogram_goodedges95} and \ref{Lisa700_histogram_goodedges90} show the number of times items are observed on good edges where a good edge is considered one above the $95$th quantile of true probabilities on edges and above the $90$th respectively.  For this network, $95^{\text{th}}$ quantile is a true probability of $0.5$ and the $90^{\text{th}}$ quantile is a true probability of $0.124$ . Figure \ref{Lisa700_realisations_goodedges95} and \ref{Lisa700_realisations_goodedges90} shows the results for $10$ repetitions of the greedy policy. }}
\end{figure}

To help understand why the BL model is giving better performance than the independent model, we look more closely at the results for the greedy heuristic for the network shown in Figure \ref{EnronNetwork_700}, with $669$ nodes and focus on the independent model with $P_{11} = 0.13$.  Figure \ref{Lisa700_cumulativerel_greedy} shows the mean cumulative number of relevant observations over $50$ runs on the greedy policy as well as the cumulative number of relevant observations for 10 realisations.  The BL model generally finds the relevant items faster than the independent model.  The cumulative number of times the algorithm changes edges is shown in Figure \ref{Lisa700_cumulativechanges_greedy} for the two models. The BL model tends to change edges less often, sticking on the same edge for longer periods of time.  Towards the end of the runs, the independent model starts sticking with edges more often.  This corresponds to an increase in the number of relevant items found; 
after a while the independent model finds edges with high relevance value.  \\

Figure \ref{Lisa700_histogram_goodedges95} and Figure \ref{Lisa700_histogram_goodedges90} show the distribution of times that each algorithm selects an edge with a true probability greater than the $95$th and $90$th quantile respectively.  For this network, those values correspond to probabilities of $0.5$ and $0.124$.  Although the algorithm with the independent model sometimes selects more edges with probabilities greater than $0.5$ than the BL model, it's performance is more varied.  Furthermore, when we look at the edges with probability above the $90$th quantile, the BL model selects a far larger number of these edges, generally picking edges above the $90$th quantile over half the time, compared to the independent model which selects these edges less than half the time.  The sharp jumps up in Figure \ref{Lisa700_realisations_goodedges95} suggests that when the BL model finds a good edge, it will stick on the good edge for a while, before moving on and finding another edge.  There are far more smaller 
jumps in the independent model; even when it finds a good edge it doesn't necessarily stick with it.  The steep gradient of Figure \ref{Lisa700_realisations_goodedges90} for the BL model suggest that when the algorithm moves off a very good edge it is likely to move to another one which is still good (above the $90$th quantile).

\section{Discussion}
\label{secDicussion}
We have considered the problem of searching a network of communications to find those which are relevant to a query.  
We show that the BL methods provide a natural approach to modelling such data.  For both simulated data and data from the Enron corpus,
the BL model gives comparable performance to the methods of \cite{dimitrov2015finding} and for networks where this method is infeasible 
BL gives superior performance to methods that ignore the network structure.  \\

Whilst we have described the BL approach for one specific model, it can easily be applied more widely. For example it is straightforward to allow for differences between participants nodes, such as those due to
covariate information, by allowing for different mean probabilities of relevance for different nodes. Similarly, we can generalise the dependence structure assumed by the BL method, for example allowing
nodes that communicate more frequently to be more strongly linked, by altering how the prior variance is specified. Furthermore, the BL approach could be applied to models which assumed a different distribution
for the probability of relevance of an item on edge given the relevance of the participants it is between. This model could include covariate information about the item, or be different depending on which of the participants
is the sender and which is the receiver. 

 \bibliography{PaperRefs}
 
 \appendix
 \section{Proof of Lemma \ref{tightconst}} \label{proof_lemma_tightconst}
 Consider two optimization problems: \eqref{constrainedBLopt} without its constraints, which is equivalent to the optimisation problem \eqref{standardBLopt}, and \eqref{constrainedBLopt} with the constraints.  Both problems are convex minimisation problems, with the same objective function, and if we assume that $\text{Var}(Z)$ is positive definite, they are strongly convex.  Suppose \eqref{constrainedBLopt} has an optimal solution $h^*$, for which none of the constraints are tight.  The same solution is the optimal for \eqref{standardBLopt} because the objective function is strongly convex.  That is, you cannot move in any direction $d$, to a new solution for \eqref{standardBLopt}, $h^* + \epsilon d$, and decrease the objective. This shows the contrapositive of the statement: If \eqref{constrainedBLopt} has an optimal solution without tight constraints, \eqref{constrainedBLopt} and \eqref{standardBLopt} have the same optimal solution.  If they do not have the same solution, then \eqref{constrainedBLopt} has 
to have a tight constraint.  
 
 \section{Proof of Lemma \ref{lemma_equalc_const}} \label{proof_lemma_equalc_const}
 The optimisation problem in \eqref{constrainedBLopt_equalc}, 
  \begin{equation*}
\begin{aligned}
& \underset{\mathbf{h}^k}{\text{minimise}}
& & E \left[ \left( Z_k -h_0^k - \sum_{i=1}^n h_i^k Y_i  \right)^2 \right] ,\\
& \text{subject to}
& &h_0^k + \sum_{i=1}^n h_i^k y_i = c,
\end{aligned}
\end{equation*}
can be solved optimally using the method of Lagrange multipliers to give an analytical solution.  Let:
\begin{align}
 \Lambda (\mathbf{h}^k,\lambda) =  E \left[ \left( Z_k - h_0^k - (\mathbf{h}_{1:n}^k)^T \mathbf{Y}  \right)^2 \right] + \lambda (h_0^k + (\mathbf{h}_{1:n}^k)^T \mathbf{y} - c), \label{lagrange_objective1}
\end{align}
where $\mathbf{h}_{1:n}^k = (h_1^k,\ldots,h_n^k)^T$.  The optimisation problem is equivalent to solving:
\begin{align}
 \underset{\mathbf{h}^k,\lambda}{\text{minimise }} \Lambda (\mathbf{h}^k,\lambda).
\end{align}
Differentiating \ref{lagrange_objective1} with respect to $h_0^k$, $\mathbf{h}_{1:n}^k$ and $\lambda$ gives:
\begin{align}
    \frac{\partial  \Lambda(h_0^k,\mathbf{h}_{1:n}^k,\lambda)}{\partial h_0^k} &= 2h_0^k - 2E[Z_k] + 2 (\mathbf{h}_{1:n}^k)^T E[\mathbf{Y}] + \lambda  \label{partial_h01}\\
 \frac{\partial \Lambda(h_0^k,\mathbf{h}_{1:n}^k,\lambda)}{\partial \lambda} &= h_0^k + (\mathbf{h}_{1:n}^k)^T \mathbf{y} - c \label{partial_lam1}\\
 \bigtriangledown_{\mathbf{h}_{1:n}^k} \Lambda(h_0^k,\mathbf{h}_{1:n}^k,\lambda) &= -2E[Z_k\mathbf{Y}] + 2h_0^kE[\mathbf{Y}]^T + 2(\mathbf{h}_{1:n}^k)^T E[\mathbf{Y}\mathbf{Y}] + \lambda \mathbf{y}^T \label{partialh1}
\end{align}
The optimal solution is found by the partial derivatives \eqref{partial_h01} - \eqref{partialh1} equal to zero,
\begin{align}
 0 &= 2h_0^k - 2E[Z_k] + (\mathbf{h}_{1:n}^k)^T E[\mathbf{Y}] + \lambda,  \label{sim_h01}\\
 0 &= h_0^k + (\mathbf{h}_{1:n}^k)^T \mathbf{y} - c ,\label{sim_lam1}\\
 0 &= -2E[Z_k\mathbf{Y}] + 2h_0^kE[\mathbf{Y}]^T + 2(\mathbf{h}_{1:n}^k)^T E[\mathbf{Y}\mathbf{Y}] + \lambda \mathbf{y}^T, \label{simh1}
\end{align}
and solving the set of simultaneous equations.  Rearranging \eqref{sim_h01} to give $h_0^k$ in terms of $\lambda$ and $\mathbf{h}_{1:n}^k$ and substituting into equations \eqref{sim_lam1} and \eqref{simh1} gives:
\begin{align}
 0 &= E[Z_k] +   (\mathbf{h}_{1:n}^k)^T (\mathbf{y} - E[\mathbf{Y}]) - \frac{\lambda}{2} - c \label{h0_inlam1}\\
 0 & = -2E[Z_k \mathbf{Y}] + 2\left( E[Z_k] - (\mathbf{h}_{1:n}^k)^T E[\mathbf{Y}] - \frac{\lambda}{2} \right)E[\mathbf{Y}]^T + 2(\mathbf{h}_{1:n}^k)^TE[\mathbf{Y}\mathbf{Y}] + \lambda \mathbf{y}^T \nonumber \\
  & = -2\text{Cov}(Z_k,\mathbf{Y}) + 2(\mathbf{h}_{1:n}^k)^T\text{Var}(\mathbf{Y}) + \lambda(\mathbf{y}-E[\mathbf{Y}])^T \label{h0_inh1}
\end{align}
Rearranging \eqref{h0_inlam1}, we get:
\begin{align}
\lambda = 2E[Z_k]+2(\mathbf{h}_{1:n}^k)^T(\mathbf{y}-E[\mathbf{Y}]) - 2c, \label{lamepression1}
\end{align}
and substituting into \eqref{h0_inh1},
\begin{align}
0 = - 2\text{Cov}(Z_k,\mathbf{Y}) + 2(\mathbf{h}_{1:n}^k)^T\text{Var}(\mathbf{Y}) + \left( 2E[Z_k]+2(\mathbf{h}_{1:n}^k)^T(\mathbf{y}-E[\mathbf{Y}]) - 2c \right)(\mathbf{y}-E[\mathbf{Y}])^T
\end{align}
Finally, rearranging we get an expression for the optimal value of $(\mathbf{h}_{1:n}^k)^T$, which satisfies \eqref{constrainedBLopt_equalc}
\begin{align}
  \left(\text{Cov}(Z_k,\mathbf{Y}) + (c - E[Z_k])(\mathbf{y}- E[\mathbf{Y}])^T\right)\left( \text{Var}(\mathbf{Y}) + (\mathbf{y} - E[\mathbf{Y}])(\mathbf{y}-E[\mathbf{Y}])^T\right)^{-1}.
\end{align}
Substituting into equation \ref{sim_lam1} gives the optimal value for $h_0^k$:
\begin{align}
 h_0^k &= c - (\mathbf{h}_{1:n}^k)^T \mathbf{y}.
\end{align}

 \section{Proof of Lemma \ref{lemmaprior}} \label{prooflemmaprior}
The expected value of $\mathbf{Y}$ can be calculated from:
\begin{align}
 E[Y_{uv}] &= E[E[Y_{uv}\mid P_{uv}]] \\
	  & = n_{uv} E[P_{uv}].
\end{align}
where $n_{uv}$ is the number of items observed on edge $(u,v)$ to date.  Similarly, the analytical formula for $E \left[ Z_k Y_{uv}  \right]$ is: 
\begin{align}
 E \left[ Z_k Y_{uv}  \right] &= E\left[ Z_k E\left[Y_{uv}\mid P_{uv}\right] \right] ,\nonumber \\
			     & = n_{uv} E\left[ Z_kP_{uv} \right]. 
\end{align}
The analytical formula for $E[\mathbf{Y}\mathbf{Y}]$ is defined in terms of diagonal and off diagonal terms of the matrix where:
\begin{align}
 E\left[Y_{uv}^2\right]& = E\left[ E \left[ Y_{uv}^2 \mid P_{uv} \right]  \right], \nonumber \\
		& = E\left[ \text{Var}( Y_{uv} \mid P_{uv} ) + E \left[ Y_{uv} \mid P_{uv} \right]^2 \right], \nonumber \\
		&=E \left[ n_{uv}(n_{uv} - 1) P_{uv}^2 + n_{uv} P_{uv} \right], \nonumber \\
		& = n_{uv}(n_{uv} - 1) E \left[ \text{Var}(P_{uv}\mid Z_u,Z_v) + E\left[ P_{uv} \mid Z_u,Z_v \right]^2 \right]+ n_{uv} E \left[P_{uv} \right],
\end{align}
and
\begin{align}
 E \left[ Y_{uv}Y_{ij} \right] &= E \left[ E \left[ Y_{uv}Y_{ij} \mid P_{uv},P_{ij} \right] \right], \nonumber  \\
			      &= E \left[ E \left[ Y_{uv}\mid P_{uv} \right] E \left[ Y_{ij}\mid P_{ij} \right] \right], \nonumber \\
			      &= n_{uv}n_{ij} E[P_{uv}P_{ij}].
\end{align}

\section{Calculations for analytical equations in Lemma \ref{lemmaprior} } \label{approx_priorvals}
Lemma \ref{lemmaprior} requires the values $E \left[\text{Var}(P_{uv}\mid Z_u,Z_v)\right]$,  $E\left[ E\left[ P_{uv} \mid Z_u,Z_v \right]^2 \right]$ and $E\left[ Z_kP_{uv} \right]$.  In order to calculate $E \left[\text{Var}(P_{uv}\mid Z_u,Z_v)\right]$, we require the joint distribution over $Z_u,Z_v$.  For two random variables these can be calculated exactly from their prior expectation and covariance, to give $P(Z_u,Z_v)$.  From this we get:
\begin{align}
 E \left[\text{Var}(P_{uv}\mid Z_u,Z_v)\right] = \sum_{Z_u,Z_v \in \left\{ 0,1 \right\}} P(z_u,z_v)\text{Var}(P_{uv}|z_u,z_v).
\end{align}
Similarly, calculating $E\left[ E\left[ P_{uv} \mid Z_u,Z_v \right]^2 \right]$ we can get $P(Z_u,Z_v)$ from the prior expectation and covariance.  From this we get:
\begin{align}
E \left[E[P_{uv}\mid Z_u,Z_v]^2\right] = \sum_{Z_u,Z_v \in \left\{ 0,1 \right\}} P(z_u,z_v)E[P_{uv}|z_u,z_v]^2.
\end{align}
Calculating $E\left[ Z_kP_{uv} \right]$ can require calculating the joint distribution over up to three $Z$'s.  This is estimated using the maximum entropy distribution as $\tilde{P}(Z_k,Z_u,Z_v)$, and using this, we get:
\begin{align}
 E\left[ Z_kP_{uv} \right] = \sum_{Z_u,Z_v,Z_k \in \left\{ 0,1 \right\}} \tilde{P}(z_k,z_u,z_v) z_k E[P_{uv}|z_u,z_v].
\end{align}

\section{Bayes UCB Algorithm} \label{BayesUCB_Alg}

\begin{algorithm}
\caption{Bayes-UCB for network based searches}
\label{BayesUCB}
\textbf{Initialize:} For $t = 1$:
\begin{enumerate}
 \item For each arm $j = 1,\ldots,K$, do:
 \begin{enumerate}
 \item Compute
 \begin{align}
  q_j(t) = \Phi^{-1}(0.5,\hat{\mu}_j^0,\hat{\sigma}_j^0)
 \end{align}
 \end{enumerate}
\item Draw arm $I_t = argmax_{j=1,\ldots,k} q_j(t)$
\item Get reward and update distributions
\end{enumerate}
\textbf{Iterate:} For $t$ = $2$ to $n$ do:
\begin{enumerate}
  \item For each arm $j = 1,\ldots,K$, do:
  \begin{enumerate}
  \item Compute:
  \begin{align}
   q_j(t) = \Phi^{-1}(1-\frac{1}{t},\hat{\mu}_j^{t-1},\hat{\sigma}_j^{t-1})
  \end{align}
  \end{enumerate}
 \item draw arm $I_t = argmax_{j=1,\ldots,k} q_j(t)$
 \item Get reward and update  distributions.
\end{enumerate}
\end{algorithm}

\end{document}